\RequirePackage{fix-cm}
\documentclass[twocolumn]{svjour3}          
\smartqed  
\usepackage{graphicx}
\usepackage[hyphens]{url}
\usepackage{forest}
\usepackage{longtable}
\usepackage{listings}
\usepackage{xcolor}
\usepackage{hyphenat}
\usepackage{hyperref}

\definecolor{codegreen}{rgb}{0,0.6,0}
\definecolor{codegray}{rgb}{0.5,0.5,0.5}
\definecolor{codepurple}{rgb}{0.58,0,0.82}
\definecolor{backcolour}{rgb}{0.95,0.95,0.92}
\lstdefinestyle{mystyle}{
    backgroundcolor=\color{backcolour},   
    commentstyle=\color{codegreen},
    keywordstyle=\color{magenta},
    numberstyle=\tiny\color{codegray},
    stringstyle=\color{codepurple},
    basicstyle=\ttfamily\footnotesize,
    breakatwhitespace=false,         
    breaklines=true,                 
    captionpos=b,                    
    keepspaces=true,                 
    numbers=left,                    
    numbersep=5pt,                  
    showspaces=false,                
    showstringspaces=false,
    showtabs=false,                  
    tabsize=2
}

\journalname{Requirements Engineering}
\begin{document}

\title{An Approach for Performance Requirements Verification and Test Environments Generation}


\author{Waleed Abdeen \and Xingru Chen \and Michael Unterkalmsteiner}


\date{The version of record of this article, first published in \emph{Requirements Engineering}, is available online at Publisher’s website: \url{https://doi.org/10.1007/s00766-022-00379-3}}

\maketitle

\begin{abstract}
\emph{Background:} Model-Based Testing (MBT) is a method that supports the design and execution of test cases by models that specify the intended behaviors of a system under test. \emph{Motivation:} While systematic literature reviews on MBT in general exist, the state-of-the-art on modeling and testing performance requirements has seen much less attention. \emph{Method:} Therefore, we conducted a systematic mapping study on model-based performance testing. Then, we studied natural language software requirements specifications in order to understand which and how performance requirements are typically specified. Since none of the identified MBT techniques supported a major benefit of modeling, namely identifying faults in requirements specifications, we developed the Performance Requirements verificatiOn and Test EnvironmentS generaTion approach (PRO-TEST). Finally, we evaluated PRO-TEST on 149 requirements specifications. \emph{Results:} We found and analyzed 57 primary studies from the systematic mapping study, and extracted 50 performance requirements models. However, those models don't achieve the goals of MBT, which are validating requirements, ensuring their testability, and generating the minimum required test cases. We analyzed 77 Software Requirements Specification (SRS) documents, extracted 149 performance requirements from those SRS, and illustrate that with PRO-TEST we can model performance requirements, find issues in those requirements and detect missing ones. We detected three not-quantifiable requirements, 43 not-quantified requirements, and 180 underspecified parameters in the 149 modeled performance requirements. Furthermore, we generated 96 test environments from those models. \emph{Conclusion:} By modeling performance requirements with PRO-TEST, we can identify issues in the requirements related to their ambiguity, measurability, and completeness. Additionally, it allows to generate parameters for test environments.

\end{abstract}

\section{Introduction}
\label{intro}
Performance aspects such as time behavior, capacity, or throughput, are essential non-functional requirements (NFR) of software products. Performance testing is the process of measuring the availability, response time, throughput, and resource utilization of a software product~\cite{molyneaux_art_2014}. The importance of software performance and relation to functional requirements is acknowledged since the 1990s~\cite{smith1993software}. A real-world example is HealthCare.gov, a "health insurance exchange website" run by the United States government, where on the launch day 99\% of people who wanted to get insurance failed to register~\cite{healthcareWiki}. Further investigations showed that no adequate performance testing was performed~\cite{performanceFailureCigniti}.

Performance-related issues can have a large impact on cost, especially if those issues are not treated early~\cite{chung_non-functional_2012,clements_coming_1997,smith_performance_2001}. Another example of a software performance issue was Pokemon Go~\cite{PokemonRolleOUt}, a  mobile game that, after the initial roll-out, became unusable in many countries. The large number of users caused server failures, leading to a delayed roll-out of the game to reduce the load~\cite{PokemonRolleOUt}. A potential reason for such a failure is the different nature of performance requirements compared to functional requirements, which makes it difficult for developers to translate performance requirements into written code~\cite{woodside_future_2007}. Therefore, performance testing is necessary, since it can detect the causes of performance-related issues and verify whether the software product meets the requirements or not~\cite{woodside_future_2007}. 

Model-based testing (MBT) is a software testing approach that uses an abstraction of the system (or part thereof) to generate test cases~\cite{utting_taxonomy_2012}. According to a software testing survey conducted in Canada~\cite{garousi_survey_2013}, more than 35\% of the respondents use MBT approaches to generate test cases in their projects. This indicates that MBT is prevalent in the industry. MBT forces testability into the product design when creating the model. The model is created from the requirements and describes the behavior of the system. Successfully modeled system requirements indicate that those requirements are testable, complete, and can be validated since they were formalized in an unambiguous manner~\cite{hasling_model_2008}.

Many studies explored the state-of-the-art of MBT~\cite{dias_neto_survey_2007,dias-neto_picture_2010,felderer_model-based_2016,haser_software_2014,utting_taxonomy_2012,utting_legeard_2006}. Utting et al.~\cite{utting_taxonomy_2012,utting_legeard_2006} created a taxonomy of existing MBT approaches and tools, and Dias-Neto et al.~\cite{dias_neto_survey_2007,dias-neto_picture_2010} systematically reviewed the literature of MBT in 2007 and 2010. These studies agreed that the existing MBT approaches focus on testing the functional rather than the non-functional part (i.e., quality aspects) of the system. Later, Häser et al.~\cite{haser_software_2014} reviewed the literature for model-based integration testing for NFR, and Felderer et al.~\cite{felderer_model-based_2016} model-based security testing. A look at the state-of-the-art for model-based performance testing is missing.

In this paper, we study the current status of model-based performance testing, and identify approaches that we can use to model different aspects of performance requirements. Then, we propose the Performance Requirements verificatiOn and Test EnvironmentS generaTion approach (PRO-TEST) which supports model-based performance testing by checking the ambiguity, measurability, and completeness of performance requirements, and generating test environments. Finally, we evaluate PRO-TEST on real software requirements specifications.

The main contributions of this study are:
\begin{enumerate}
    \item A categorization of MBT studies in the context of performance requirements, based on the performance aspect, testing level, study type, research method, model type, application type, and contribution.
    \item A categorization of the Software Requirements Specifications (SRS) from a public repository~\cite{ferrari_towards_2017}, based on the described application type and performance requirements.
    \item PRO-TEST, an approach to model performance requirements to verify them, understand what should be tested, and generate test environments.
    \item An evaluation of PRO-TEST, illustrating its benefits and drawbacks.
\end{enumerate}   

The remainder of this paper is organized as follows. Section~\ref{sec:bgrw} introduces the concepts of software performance and model-based testing, and reviews related work. Section~\ref{sec:rm} illustrates the design and methodology used in our research and the validity threats. In Section~\ref{chap:pr} we present state-of-the-art and state-of-practice of model-based performance testing. Section~\ref{sec:pro-test} presents PRO-TEST and the obtained benefits but also the faced challenges when modeling performance requirements. We discuss PRO-TEST in relation to literature in Section~\ref{chap:pro-test discuss}. Section~\ref{sec:arq} answers our research questions. Finally, we conclude the paper in Section~\ref{sec:cfw} with directions for future work.

\section{Background and related work}\label{sec:bgrw}
In this section, we briefly review aspects of software performance, model-based testing, and related work.

\begin{table*}[htb]
\caption{Quality models and their related performance aspects}
\label{tab:quality model and performance aspect}
    \begin{tabular}{p{0.20\textwidth}p{0.75\textwidth}}
        \hline\noalign{\smallskip}
        \textbf{Quality Model Name} &  \textbf{Performance Aspect} \\
        \noalign{\smallskip}\hline\noalign{\smallskip}
        McCall's & Execution Efficiency, 
        Storage Efficiency 
        \\
        Bohem's  & Accountability, Device Efficiency,
        Accessibility 
        \\
        Dromey’s  & Internal Efficiency, Descriptive Efficiency
        \\
        FURPS & Speed, Efficiency, Availability, Accuracy, Throughput, Response Time, Recovery Time, Resource Usage
        \\
        ISO9126  & Time Behavior, Resource Utilization, Efficiency Compliance 
        \\
        ISO25010  & Time Behavior, Resource Utilization, Capacity
        \\
        \noalign{\smallskip}\hline
    \end{tabular}
\end{table*}

\subsection{Software performance}
\label{sec:sw-performance}

Software performance is considered in many software quality models~\cite{al-qutaish_quality_2010,khosravi_quality_2004}. Synthesizing these quality models, as shown in Table~\ref{tab:quality model and performance aspect}, the main aspects of software performance are time behavior~\cite{coallier2001software,grady_software_1987,ISO25010}, capacity~\cite{ISO25010}, resource utilization~\cite{coallier2001software,grady_software_1987,ISO25010}, speed/throughput\footnote{The meaning of the symbol "/" is "or". We kept both words because they are both used frequently in performance.} \cite{grady_software_1987} and efficiency~\cite{boehm_verifying_1984,coallier2001software,dromey1995model,mccall1977factors}. Next, we provide a definition of these software performance aspects.

\paragraph{Time Behavior:} the time required to perform specific tasks or complete requests. It usually has multiple instances or values depending on different anticipated capacities (i.e., the number of users). This aspect is included in all three models (ISO9126, ISO25010, and FURPS) as time behavior or response time. It is an explicit aspect, that is used by the users to infer software performance. It could have a direct effect on the usability of the software.

\paragraph{Resource Utilization:} the amount or percentage of the resources used to run the software. The software should not always utilize all resources when running, instead, it should be limited to a specific amount so that it has a margin for peak times and new updates that would require more resources.

\paragraph{Capacity:} the maximum capacity (in terms of requests, sessions, users, data, etc.) that the system can handle without crashing. This aspect is crucial when planning for the project in later stages, especially when considering scalability. If not accounted for, it could result in an overload of the system, which would affect the business operations and lead to extra charges. Capacity gives an insight into the anticipated data size used by the software, which would affect the decision regarding the required resources for the system to operate.

\paragraph{Speed/Throughput:} the number of requests or processes per time unit that the system can handle while still maintaining the time behavior requirements.

\paragraph{Efficiency:} the relation between the output (i.e., time behavior, speed) and the input (i.e., capacity, resource utilization). This is a relatively complex aspect since it is affected by all other mentioned aspects of performance. 

\subsection{Model-Based Testing}
\label{sec:MBT}

Model-based testing is a software testing technique that automates the process of test case generation from a model that represents the system under test (SUT). MBT consists of three main tasks~\cite{schieferdecker_model-based_2012}: designing a functional test model, determining test generation criteria, and generating the tests. The model could be an end-to-end model, e.g., a business process or per function process model. Abstract test cases are generated from a systems' model by random generation, search-based algorithms, model–checking, symbolic execution, theorem proving, or constraint solving~\cite{utting_taxonomy_2012,li_survey_2018}. Then a tool builds the test skeleton to test the software. 

\begin{figure}[htb]
\centering
    \includegraphics[width=0.9\columnwidth]{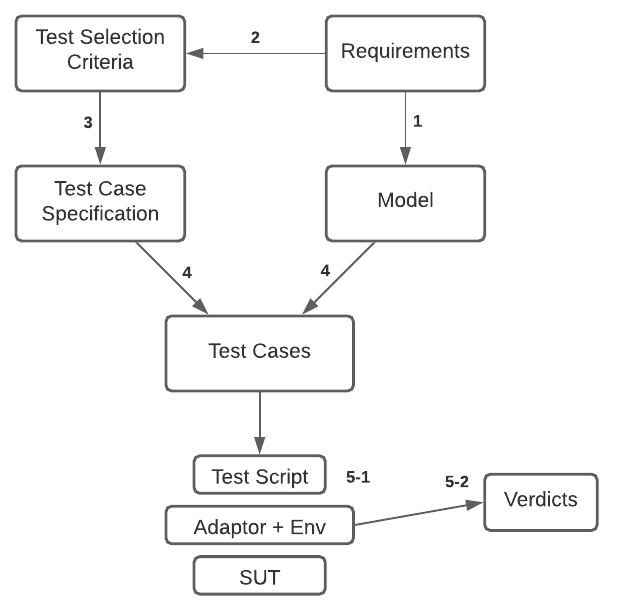}
\caption{MBT Process Diagram from Utting et al.~\cite{utting_taxonomy_2012}}
\label{fig:mbt-process-utting}
\end{figure}

Utting et al.~\cite{utting_taxonomy_2012} present five steps of the MBT process. We illustrate this process in Figure~\ref{fig:mbt-process-utting}. In \textit{Step~1}, a test model is created from the requirements. The model can be either created specifically for testing or reuse some parts of the models used at the design phase. In the case of the latter, the test model should be independent of the design model, so issues in the design phase do not appear in the test model. A model should be verified with little effort to ensure the efficiency of the MBT approach. In \textit{Step~2}, test selection criteria are defined, which will set the rules for the automatic generation of test cases. Examples of test selection criteria are system functionality (requirements-based), the structure of the test model, or properties of the environment. In \textit{Step~3}, test case specifications are written as a more formal representation of test case selection criteria. In \textit{Step~4}, the test specification is, with help of the model, transformed into concrete test cases. At this stage, algorithms are used to select the minimum set of test cases that ensure full test coverage. In \textit{Step~5}, the tests are run on the SUT in a test environment. First, test inputs are fed to the function under test (5-1), then the test verdicts are recorded by comparing the test results with the expected outcome (5-2).

There are many benefits associated with MBT. It has been shown to be effective in testing real-time adaptive systems~\cite{abdelgawad_model-based_2017}, verifying the system behavior, and identifying possible performance enhancements. Furthermore, the benefits of MBT automation are generally more numerous the more testing the system requires~\cite{pretschner2005one}. Another benefit is that MBT finds missing and unclear requirements by modeling the systems' requirements~\cite{myers_art_2004,paradkar_specificationbased_1997}. Besides, MBT can make the requirements more understandable for software engineers~\cite{woodside_future_2007}. Since performance requirements are often written at a high abstraction level, it may be difficult to understand how they impact software design and code. This could be made easier by modeling functional and non-functional requirements using the same model. A UML activity diagram that models functional requirements could be annotated with performance requirements~\cite{da2011generation}. We could see in the resulting model where the performance requirements apply in the software.

\subsection{Related work}
There exist many studies that investigate MBT to test functional requirements, while fewer studies focus on non-functional requirements. Utting et al. created in 2006~\cite{utting_legeard_2006} and 2012~\cite{utting_taxonomy_2012} respectively a taxonomy for model-based testing to categorize the existing approaches and tools, as well as to classify their usefulness. Their study focused on functional requirements testing. In general, there is no clear distinction between functional and non-functional requirements when MBT is applied~\cite{hooda2013future}. 

Although MBT for non-functional requirements is not explored extensively, there are still some studies in this area. A systematic review (78 papers) of MBT approaches by Dias-Neto et al.~\cite{dias_neto_survey_2007}, published in 2007, was not limited to functional requirements and explores the non-functional aspects considered by the models. Some limitations of using MBT for non-functional requirements were pointed out by the study. The irregular behavior of software users makes it hard to create a behavioral model of non-functional requirements. Another challenge is the limited support for non-functional aspects in the existing MBT approaches; NFRs like usability, reliability, and security were not supported. Moreover, the majority of MBT approaches proposed by research are never used in industry~\cite{dias_neto_survey_2007}. 

Dias-Neto's original study was renewed in 2010~\cite{dias-neto_picture_2010} (including 219 papers), with a focus on the techniques used for modeling, coverage, and the challenges of MBT. This study introduces selection criteria for MBT approaches based on their characteristics. The use of MBT techniques was still difficult, as observed in their previous study in 2007. Apparently, NFRs (usability, reliability, and security) that were not possible to test with MBT (according to the 2007 review~\cite{dias_neto_survey_2007}), started to get some attention in research. The difference between these studies~\cite{dias_neto_survey_2007,dias-neto_picture_2010} and the systematic mapping study presented in Section~\ref{chap:pr} is that ours has a more narrow focus on model-based performance testing.

In 2014, Häser et al.~\cite{haser_software_2014} conducted a systematic literature review of model-based integration testing. They asked in their research questions about the software paradigms, assessment type, and which NFR can be tested with MBT approaches. However, they did not ask whether the MBT approach tests different aspects of an NFR (i.e., what aspects of performance were tested using these MBT approaches?), and they scoped their research to integration testing. Their findings indicate a lack of research in model-based integration testing for NFR.  

In 2016, Felderer et al.~\cite{felderer_model-based_2016} presented a taxonomy and systematic classification of model-based security testing. They extended the study of Dias-Neto et al.~\cite{dias-neto_picture_2010} while focusing on security requirements. Woodside et al.~\cite{woodside_future_2007} described the domain of software performance engineering (SPE). They did a survey of current work on a sample of papers in SPE and pictured the future of SPE. They collected some models and methods which are used for performance and listed many benefits of modeling performance. The focus of that study is to provide a look at the future of model-based performance testing. In contrast, our study focuses on identifying current techniques that can be used in practice.

Motivated by this research gap, the lack of systematic reviews in MBT of performance requirements, the limited support for NFR in general, and performance in particular in existing techniques, we focus our research on finding and studying different performance requirements models, for the purpose of using them in MBT.

\section{Research methodology}\label{sec:rm}

To achieve our research aim defined in Section~\ref{intro}, we have identified the following four objectives.

\begin{itemize}
    \item \textbf{O1} Identify which aspects of performance are important and can be modeled.
    \item \textbf{O2} Identify modeling techniques and methods that suit performance requirements.
    \item \textbf{O3} Identify a modeling approach that can validate performance requirements, ensure that those requirements are testable, and support the generation of test cases, all three of which are key aspects of MBT.
    \item \textbf{O4} Evaluate the identified modeling approach on a set of requirements specifications.
\end{itemize}

In alignment with those objectives, we define our research questions in Table~\ref{tab:research-questions}.

\begin{table*}[htb]
\centering
\caption{Research Questions}
\label{tab:research-questions} 
    \begin{tabular}{p{0.075\textwidth}p{0.325\textwidth}p{0.40\textwidth}p{0.10\textwidth}}
    \hline\noalign{\smallskip}
    Number & Research Question & Purpose & Objective \\
    \noalign{\smallskip}\hline\noalign{\smallskip}
         RQ1 & Which aspects of performance requirements are used in MBT? &
         There are many performance aspects, e.g., time, speed, and capacity, as explained in Section \ref{sec:sw-performance}. Those aspects may have different ways of modeling and testing. & O1
         \\
         RQ1.1 & Which aspects of performance requirements have been studied? &
         Explore the studied aspects of performance requirements in MBT. & O1
         \\
         RQ1.2 & Which aspects of performance requirements can be modeled?
         &
         Explore the usage of MBT to model different aspects of performance requirements. & O1
         \\
         RQ1.3 & Which aspects of performance requirements are used in real-life projects?
         &
         Explore the performance aspects that are specified and relevant in real-life projects. & O1
         \\
         RQ2 & How to implement MBT on performance requirements aspects?
         &
         Explore the different MBT approaches that support the modeling of performance requirements to understand the current state of the art of MBT for performance requirements. & O2, O3
         \\
         RQ2.1 & What type of models can be used to model performance requirements aspects?
         &
         There are many models used in MBT. However, that does not mean all of them could be used to model all aspects of performance requirements. & O2
         \\
         RQ2.2 & Which performance requirements models achieve the goals of MBT?
         &
         Find models that achieve MBT goals, which are validating requirements, ensuring their testability and generating the minimum required test cases. & O3
         \\
         RQ3 & To what extend is the identified approach effective at modeling performance requirements written for real-life projects?
         &
         Evaluate the modeling approach that we identified in the previous step, to ensure its applicability on real-life projects with different aspects of performance requirements. & O4
         \\
    \noalign{\smallskip}\hline
    \end{tabular}
\end{table*}

\begin{figure}[htb]
\centering
    \includegraphics[width=0.9\columnwidth]{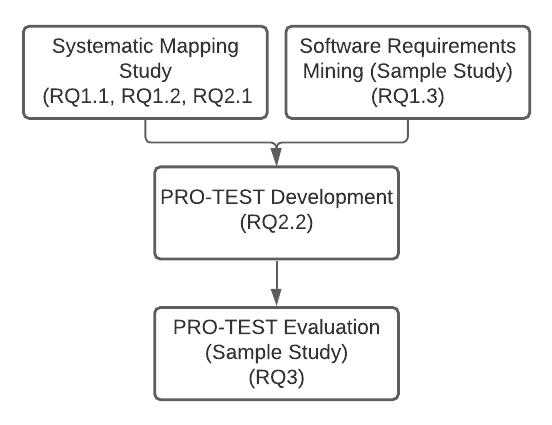}
\caption{Research Methodologies}
\label{fig:res-meth}
\end{figure}

Figure~\ref{fig:res-meth} shows the steps of our research in alignment with the research questions. First, we start with a systematic mapping study (SMS). The mapping study is an appropriate method for gaining an overview of a particular research area. We explored which performance aspects were studied and modeled using MBT (RQ1,1, RQ1.2), and what models exist to model performance requirements (RQ2.1). Second, we conducted a sample study on real-project requirements, for the purpose of finding out the relevance of performance aspects in practice (RQ1.3). Based on the results from the SMS and software requirement mining, we developed PRO-TEST (RQ2.2). Finally, we conducted a sample study, to evaluate PRO-TEST (RQ3). We focus our study on the domain of software-intensive systems.

A Systematic Literature Review (SLR) and an SMS are research methodologies that systemically survey the literature but differ in their aim, execution, and outcome~\cite{kitchenham_value_2010,petersen_guidelines_2015}. An SLR aims to aggregate data from the literature and has specific research questions for that purpose, while an SMS aims to explore trends and identify gaps in research. In terms of execution, an SLR requires a quality assessment to be conducted on the extracted papers, while it is not the case for an SMS. The output of an SLR is a synthesis of the reviewed studies, while an SMS classifies a set of papers based on different dimensions. 

A sample study is a form of research done on a sample of the population for generalization~\cite{stol_abc_2018}. The data could be collected using interviews, questionnaires, metric reports, or available for access online, e.g., in a software repository.
One of the research methods associated with sample studies is software repository mining~\cite{stol_abc_2018}. Software repository mining research usually uses open-source software repositories. There is no human to collect data from, i.e., no interviews or questionnaires are involved.

The purpose of evaluating PRO-TEST is to validate that it works in practice, i.e., it can model the performance requirements and generate test environments. Similar to the software requirements mining approach described in Section~\ref{sec:software requirements mining}, we conduct again a sample study, i.e., we use an openly accessible resource for software requirements specifications.

\subsection{Systematic mapping study}
\label{sec:sms}

We developed the SMS protocol based on the SLR conducted by Dias Neto et. al.~\cite{dias-neto_picture_2010}, following the guidelines by Petersen et al.~\cite{petersen_guidelines_2015}. There were two reasons for choosing this study by Dias-Neto. First, the research group has conducted two SLRs~\cite{dias_neto_survey_2007,dias-neto_picture_2010} on MBT using the same protocol. This provides some evidence for the repeatability of their study. Second, there was enough information presented about the search keywords and procedure, making it easier to adapt and extend the protocol. The choice of reusing and extending an existing protocol has however also disadvantages. The study of performance requirements concerns research beyond MBT, such as requirements engineering and software testing in general, software performance engineering and agile software development. Hence, we emphasize that our review covers the area of performance requirements within the scope of MBT only.

\subsubsection{Study identification}
\label{sec:sms-study identification}

\paragraph{Choosing the search strategy:} We used keyword search in digital databases similar to the search method used by Dias Neto et al. 2010~\cite{dias-neto_picture_2010}. They used six databases for their search. Two of the databases (i.e., Compendex IE and INSPEC) we did not have access to. Therefore we ran the search on the other four databases (SCOPUS, ACM, IEEE Xplore, and Web of Science). We searched the title, keywords, and abstract of the paper on SCOPUS, WoS, and ACM, while we searched the full text of IEEE (due to a limitation of the database).

\paragraph{Developing the search:} We took the search string used by Dias Neto et al.~\cite{dias-neto_picture_2010} and extended it to fit the purpose of our research. The keywords we added are related to performance. We extracted those keywords from the quality models for software performance discussed in Section~\ref{sec:sw-performance}. 
Table~\ref{tab:search-string} shows the borrowed search string and the extension with performance-related keywords.

\begin{table*}[htb]
\caption{Search strings used in the SMS}
\label{tab:search-string}
\centering
    \begin{tabular}{p{0.20\textwidth}p{0.75\textwidth}}
        \hline\noalign{\smallskip}
        Description & Keywords 
        \\
        \noalign{\smallskip}\hline\noalign{\smallskip}
        Borrowed search string from Dias Neto et al. & 
        (("model based test") OR ("model based testing") OR ("model driven test") OR ("model driven testing") OR ("specification based test") OR ("specification based testing") OR ("specification driven test") OR ("specification driven testing") OR ("use case based test") OR ("use case based testing") OR ("use case driven test") OR ("use case driven testing") OR ("uml based test") OR ("uml based testing") OR ("uml driven test") OR ("uml driven testing") OR ("requirement based test") OR ("requirement based testing") OR ("requirement driven test") OR ("requirement driven testing") OR ("finite state machine based test") OR ("finite state machine based testing") OR ("finite state machine driven test") OR ("finite state machine driven testing")) AND (software)
        \\
        \noalign{\smallskip}
        Extension &
        AND (performance OR efficiency OR capacity OR load OR  speed OR responsiveness OR  stability OR timing OR ("time behaviour") OR ("time behavior") OR ("response time") OR ("response-time") OR ("resource utilization") OR ("resources utilization") OR ("resource consumption") OR ("resources consumption") OR thruput OR throughput OR spike OR stress OR volume OR size OR scalability OR peak OR ("wait time") OR latency OR delay OR workload OR ("concurrent users") OR ("concurrent requests"))
        \\
        \noalign{\smallskip}\hline
    \end{tabular}{}
\end{table*}{}

\paragraph{Evaluating the search string:}
We evaluated the quality of the search string to mitigate the risk of missing key papers. We did that in two steps:
\begin{itemize}
    \item
    We ran Dias Neto et al.~\cite{dias-neto_picture_2010} search string on the selected databases and randomly checked whether the returned research papers were presented by Dias Neto et al. in their study~\cite{dias-neto_picture_2010}.
    \item
    To validate the whole search string including the extension, we reviewed the papers published at the \textit{International Conference On Software Testing Verification And Validation (ICST)} over the period 2014-2018. We read the title and abstract to see if the topic is related to model-based performance testing. We collected the papers related to our topic and looked for them in our search results. We found three papers in the ICST conference proceedings that were not returned by our search string. After further analysis of the search string, we removed a part of Dias Neto et al. search string~\textit{(approach  OR  method  OR  methodology  OR  technique)} and adjusted our extension to ensure those papers are included.
\end{itemize}{}

\subsubsection{Selection criteria}
\label{sec:sms-selection-criteria}
We developed the following inclusion and exclusion criteria.

\paragraph{Inclusion:}
\begin{enumerate}
    \item The publication is available in full text.
    \item The publication language is English.
    \item The date of the publication is within the range of August 2009 (the date when Dias Neto et al.~\cite{dias-neto_picture_2010} conducted their search) and February 2019 (when we conducted our search).
    \item The publication proposes and/or evaluates model-based performance testing techniques.
\end{enumerate}

\paragraph{Exclusion:}
\begin{enumerate}
    \item The publication presents secondary studies, i.e., SMS, SLR, literature survey.
    \item The publication is not related to the topic model-based performance testing.
    \item Duplicated publications that refer to the same study.
    \item The publication is about model-based mutation testing
    \item Proceeding, table of content, book, tutorial, demo, editorial
\end{enumerate}

After careful analysis of the model-based mutation testing approach, we have decided to exclude it. Although it uses MBT as a basis, it is concerned with introducing faults during the test to find issues in the system rather than the modeling and test case generation.

\subsubsection{Quality assessment}
No detailed quality assessment was conducted. Since the goal of our SMS was to find a method that we can use, there was no need to evaluate the quality of each paper selected for our research.

\subsubsection{Data extraction}
\label{sec:SMS data extraction}
We extracted the following data from our and Dias-Neto et el.'s~\cite{dias-neto_picture_2010} primary studies (after we applied our inclusion/exclusion criteria).

\paragraph{Performance aspect:} 
We extracted data related to the five performance aspects discussed in Section~\ref{sec:sw-performance}, i.e., time behavior, resource utilization, capacity, throughput, and efficiency. We added a "not specified" category for those papers that do not mention or focus on a specific aspect of performance. This classification supports answering RQ1, RQ1.1, and RQ1.2.
    
\paragraph{Testing level:} Testing can be conducted on five different levels~\cite{ammann_introduction_2016}: acceptance, system, integration, module, and unit level. This classification supports answering RQ2 and determines on which level performance testing is conducted.
    
\paragraph{Study type:} 
We used Stol et al.~\cite{stol_abc_2018} to classify study types in software engineering: field study, field experiment, experimental simulation, laboratory experiment, judgment study, sample studies, formal theory, and computer simulation. This classification helps us to understand how mature the studied MBT techniques are, i.e., whether they are empirically studied and adopted by industry or initial proposals that require more empirical evidence. This is an additional criterion for choosing the model and answering RQ2, RQ2.1, and RQ2.2.
    
\paragraph{Research method:} 
The research method differs from the study type. A research method defines the set of rules and practices to follow, having a specific goal in mind, i.e., answering a set of research questions. The study type is a grouping of different research methods based on their "metaphor, purpose and goals"~\cite{stol_abc_2018}.
In software engineering research, many research methods can be associated to study types~\cite{stol_abc_2018}. Some of those methods are case study, experiment, survey, and concept development. Since there is no complete list of those research methods, we kept this classification dynamic and extracted the options directly from the research papers. This classification helps to distinguish between papers that present a new approach or theory to others that empirically evaluate existing approaches.
    
\paragraph{Model type:} 
We classified each paper based on the approach used to model performance requirements. The classification is based on the essence of the model, i.e., some models were novel while others were extensions of previous models. For example, Maâlej et al.~\cite{maalej_model-based_2012} present timed-automata, while Abbors et al.~\cite{abbors_model-based_2013} present a probabilistic extension of timed-automata. This helps to determine the frequency of the models used for performance requirements and answer RQ2.1. We did not have predetermined options for this classification, since one of our research objectives was to identify all possible modeling approaches.
    
\paragraph{Application type:}
We classify the type of the application (e.g., web application, mobile, desktop) to understand where model-based performance testing is used or studied. This is also a dynamic classification with no predetermined options.
    
\paragraph{Contribution:}
This classification assigns papers into categories based on their contribution to the field (e.g., tool, method, evaluation). With this classification, we can understand the maturity of the models.

\subsubsection{Data analysis}
We use the frequency of the extracted data, discussed in the previous section, to analyze the state-of-art in model-based performance testing.

Also, to identify a model that can achieve the MBT's goals, we examined the following aspects of the identified MBT techniques:
\begin{itemize}
    \item reported benefits of modeling performance requirements
    \item modeled performance aspects
    \item type and strength of evaluation of the proposed method
\end{itemize}

\subsection{Software requirements mining}
\label{sec:software requirements mining}

The research questions RQ1 and RQ1.3 in Table~\ref{tab:research-questions} were answered by conducting software requirements mining.
Ferrari et al.~\cite{ferrari_towards_2017} published a data set~\cite{requirementsDataset} that contains a collection of software requirements specifications gathered from various industries and applications. There are 77 SRSs in total in the collection from which we constructed a subset as described next.

\subsubsection{Selection criteria}
\label{sec:software requirements mining-Desing-selection}

Inclusion: the SRS and the individual requirements that are classified and shown in our results have the following properties.
\begin{itemize}
    \item SRS: have at least one performance requirement.
    \item Requirement: fits in one of the descriptions for performance aspects in Section~\ref{sec:sw-performance}.
\end{itemize}
Exclusion: the SRS and the individual requirements that we excluded from our classification and the results have the following properties.
\begin{itemize}
    \item SRS: without any performance requirements or not written for a software product.
    \item Requirement: does not fit in any of the performance aspects descriptions.
\end{itemize}

\subsubsection{Coding}
\label{sec:software requirements mining-classification}
Since the data in the SRS documents is of qualitative nature, we used coding to efficiently identify and extract relevant information. The codes we created are based on having three dimensions (performance aspect, application type, and quantifiability) that we describe next.

\paragraph{Performance aspect:} 
We extract five performance aspects, i.e., time behavior, resource utilization, capacity, speed/throughput, efficiency, and a general option for the performance requirements that did not fit in any of the five aspects' descriptions. We apply this classification to each performance requirement and provide thereby data to answer RQ1.3.

\paragraph{Application type:}
Similar to the SMS, we extract the type of application specified in the SRS, e.g., web application, mobile application, embedded system, etc. This allows us to evaluate whether the SRS data set is a good presentation of the population (i.e., software products).

\paragraph{Requirements quantifiability:}
Testability is one of the major criteria in requirements verification and validation~\cite{boehm_verifying_1984}. The requirement "must be specific, unambiguous, and quantitative wherever possible" such that a developer can write software code that satisfies the requirements.
The performance requirement should be quantitative and quantified to be testable. We evaluated each requirement by looking for numerical values. 

\subsection{Evaluating PRO-TEST}
\label{sec:evaluating-PRO-TEST}

We evaluated PRO-TEST on a set of realistic software requirements specifications (SRS) containing performance requirements. The evaluation was done by modeling the performance requirements and assessing the quantifiability and degree of quantification of the specified requirements, and identifying the possible missing requirements.

\subsection{Threats to validity}
In the SMS there were threats related to the data extraction methods. 1) We may have missed some papers because two databases used by Dias Neto et al.~\cite{dias-neto_picture_2010} we did not have access to. To keep this to a minimum we made sure that we use the SCOPUS database, which includes publications from different technical publishers. 2) We may have excluded papers by our search string. We extended the search string from Dias Neto et al.~\cite{dias-neto_picture_2010} study with words related to performance. This could lead to fewer results if some keywords are missing from the search string. We tried to include as many keywords as possible and used performance checklists to make sure this threat is kept to a minimum. 3) Another type of threat is related to the human factor; we could have interpreted the data in the wrong way or placed a paper in the wrong classification. We addressed this threat by having the selection and classification done by two researchers independently and the results were then compared. When conflicts were discovered the corresponding paper was discussed by both researchers and if still no consensus could be achieved, a third researcher was consulted.

In software requirements mining, the human factor also introduces threats to validity. First, we could have coded some requirements wrongly or missed out on some performance requirements from the SRS documents. We mitigated this threat by having two researchers involved in coding. The researchers coded a sample of seven SRS documents independently and compared the results. When conflicts were discovered, the corresponding requirement was discussed by both researchers. Then we divided the work equally between the two researchers. When no consensus could be achieved by the two researchers a third researcher was consulted. Second, the sample size may not be enough for generalization since the SRS collection had 77 documents that might not cover all application types or represent the population, i.e., software products.

Finally, in the implementation of PRO-TEST, the small sample size is not enough to generalize the competence of the approach. Only 34 SRS documents of the SRS collection had performance requirements, which might lead to threes issues: 1) The sample we chose might be small to represent the population, i.e., software products. 2) The SRS collection from Ferrari et al.'s study~\cite{ferrari_towards_2017} might not be a good representation for the population as well. 3) The most recent SRS document goes back to 2010, which could be considered old. A validation of the model on more recent SRS documents is required.

\section{Model-based performance testing}\label{chap:pr}
This section reports on the results from the SMS on model-based performance testing (Section~\ref{sec:state of art}) and on the prevalence of performance requirements in a publicly available repository of software requirements specifications (Section~\ref{sec:state of practice}). We discuss our findings in Section~\ref{sec:performance requirements - discussion}. 

\subsection{State of the art}
\label{sec:state of art}

\begin{figure*}[htb]
\centering
    \includegraphics{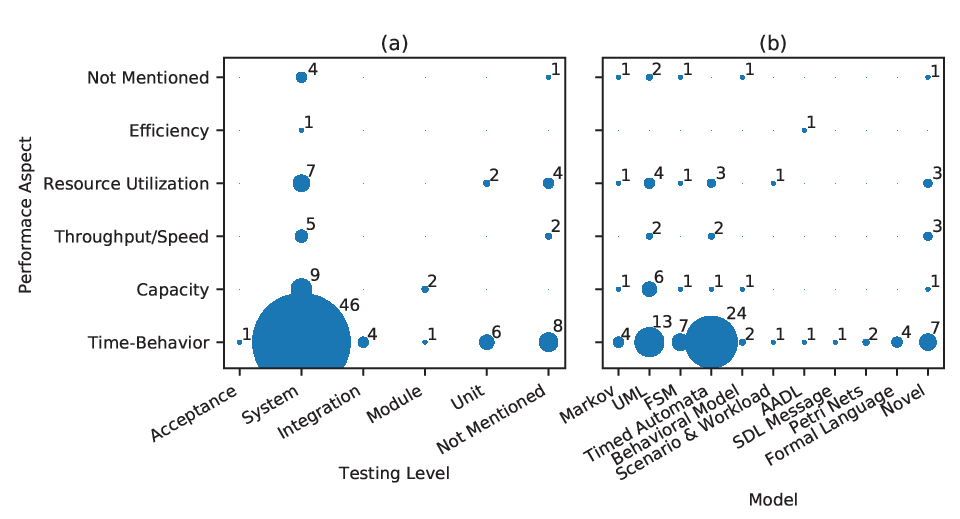}
\caption{Papers mapping between a) performance aspect and testing level b) performance aspect and model}
\label{fig:sms performance testing model}
\end{figure*}

\begin{figure*}[htb]
\centering
    \includegraphics{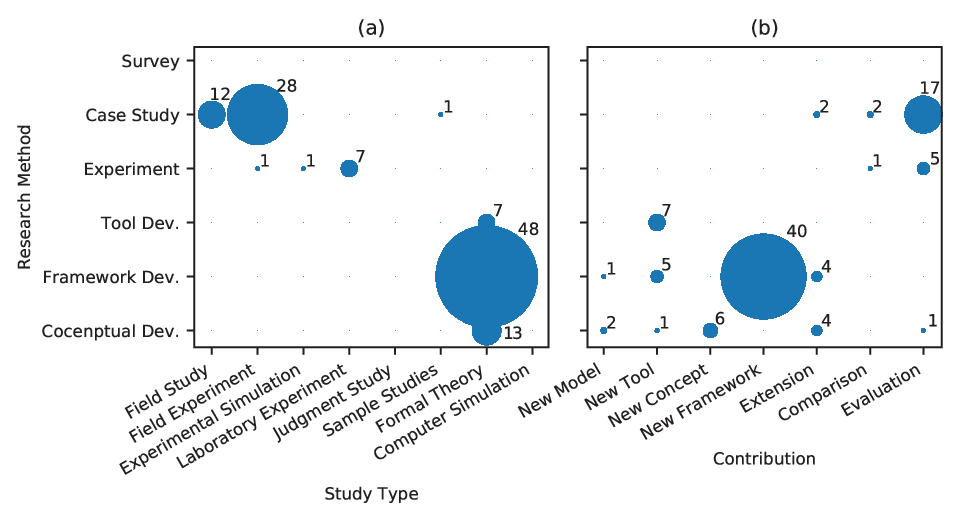}
\caption{Papers mapping between a) research method and study type b) research method and contribution}
\label{fig:sms type method contribution}
\end{figure*}

\begin{figure*}[htb]
\centering
    \includegraphics[width=0.85\textwidth]{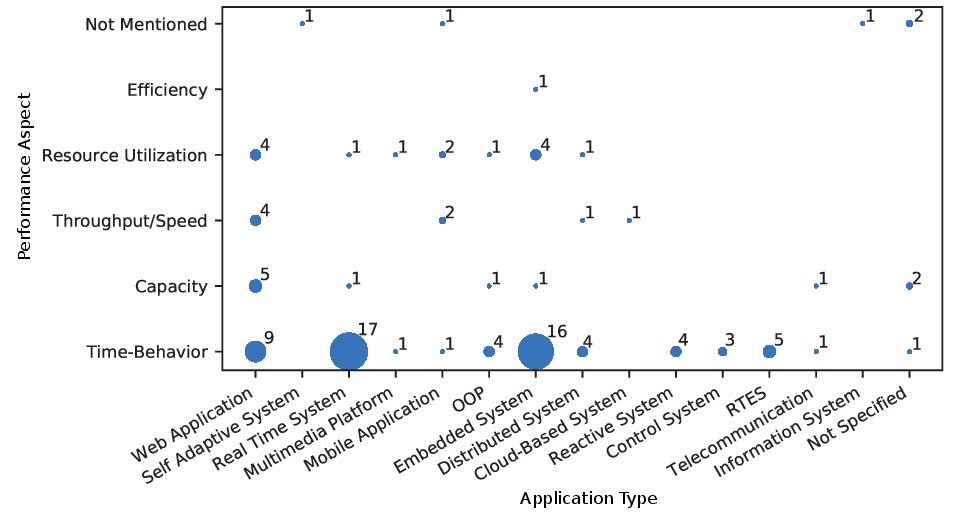}
\caption{Papers mapping between performance aspect and application type}
\label{fig:sms performance aspect and application}
\end{figure*}

We identified 57 primary studies through our database search and extracted 20 from Dias-Neto's study (see Appendix~\ref{app:included-papers-in-sms}\footnote{Additional materials including the list of primary studies, the mapping of papers to each classification, grouping of the models, data from Dias Neto's study~\cite{dias-neto_picture_2010}, the SRS collection, extracted performance requirements, the modeling of those requirements using PRO-TEST and the excluded performance requirements are available online~\cite{additionalData}.}).  A paper could be mapped to more than one value in each classification, which depends on the content of the paper. The choice of these maps was driven by our research questions.  We show in Figure~\ref{fig:sms performance aspect and application} and Figure~\ref{fig:sms performance testing model} the relation of the performance aspect with all other research area classifications. Moreover, a typical SMS should classify papers in both 1) the research area and 2) the research type~\cite{petersen_systematic_2008}, hence our choice of Figure~\ref{fig:sms type method contribution}.

In Figure~\ref{fig:sms performance testing model} the y-axis represents the performance aspects, while the x-axis in Figure~\ref{fig:sms performance testing model} (a) represents the testing level and in Figure~\ref{fig:sms performance testing model} (b) the model types. The "Not mentioned" option in the performance aspects, represents the papers that did not mention or focus on any aspect. We categorized the extracted models based on the essence of the model.

In Figure~\ref{fig:sms type method contribution} the y-axis represents the research method while the x-axis in Figure~\ref{fig:sms type method contribution} (a) represents the study type and in Figure~\ref{fig:sms type method contribution} (b) the contribution of the paper. The study type is based on the classification in Section~\ref{sec:SMS data extraction}.

In Figure~\ref{fig:sms performance aspect and application}, the y-axis represents the performance aspect while the x-axis represents the application type (grouped). We grouped the applications based on the category, purpose, and platform, e.g., web application, mobile, and embedded system. 

\begin{figure*}[htb]
\centering
    \includegraphics[width=1\textwidth]{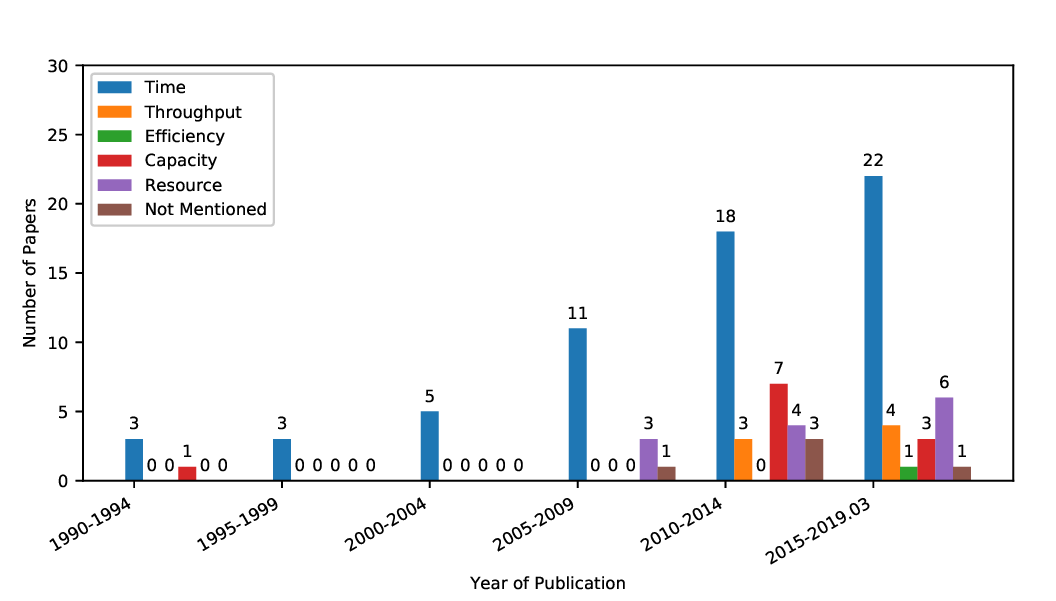}
\caption{Number of publications between 1990 and 2019 - 03}
\label{fig:sms studies per year}
\end{figure*}

Figure~\ref{fig:sms studies per year} represents the number of publications related to the topic model-based performance testing. The Figures ~\ref{fig:sms performance testing model}, \ref{fig:sms type method contribution},~\ref{fig:sms performance aspect and application} and \ref{fig:sms studies per year} are based on the results of Dias Neto et al.~\cite{dias-neto_picture_2010} (for the period 1990-2009) and our research (for the period 2009-2019). We combined the results from the two mentioned studies and present the combined results in these figures.

\subsection{Performance requirements in SRS documents}
\label{sec:state of practice}

The SRS collection contained 77 SRS documents; 34 documents contained at least one performance requirement, and 43 documents specified no performance requirements.

\begin{figure*}[htb]
\centering
    \includegraphics{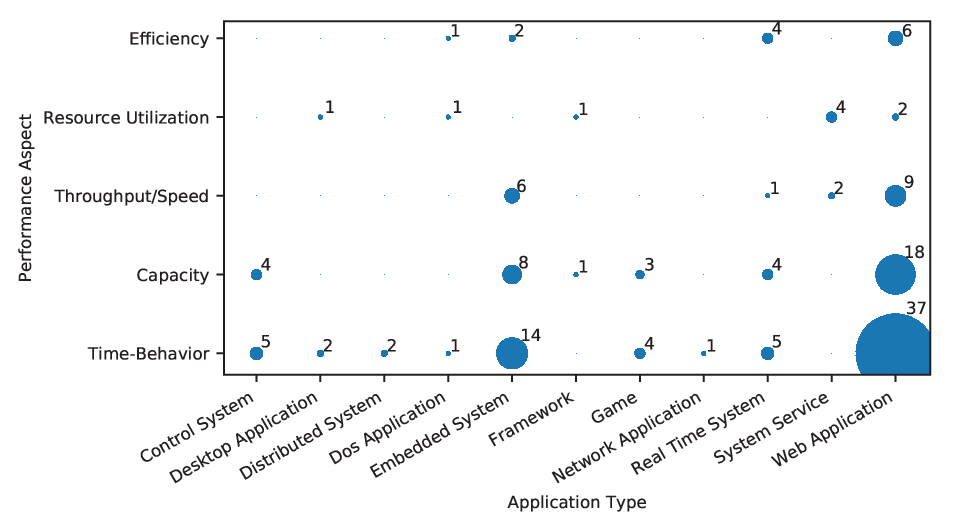}
\caption{Mapping of extracted requirements between performance aspect and application type}
\label{fig:srs performance application}
\end{figure*}

Figure~\ref{fig:srs performance application} shows the mapping of the extracted performance requirements from the SRS collection. The mapping has two dimensions, representing the performance aspect that the requirement belongs to and the application type specified in the SRS document.

\begin{figure}[htb]
\centering
\includegraphics{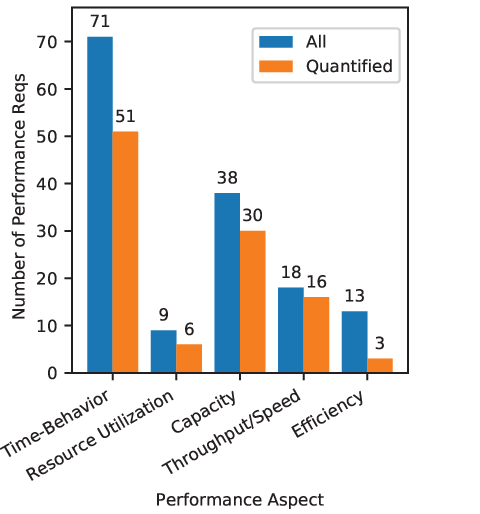}
\caption{The frequency of total and quantified performance requirements per performance aspect}
\label{fig:srs performance with all reqs and testable}
\end{figure}

To extract the performance requirements we applied the coding described in Section~\ref{sec:software requirements mining-classification}. The total number of quantifiable performance requirements was 149. However, only 106 requirements were actually quantified, thus could be modeled and tested. Figure~\ref{fig:srs performance with all reqs and testable} shows the number of extracted performance requirements per performance aspect and the quantified requirements per aspect.

\subsection{Discussion}
\label{sec:performance requirements - discussion}

Research on model-based performance testing has gained momentum over the past 30 years (Figure~\ref{fig:sms studies per year}).

Performance aspects were studied to a different extent. By far the most prevalent performance aspect in studies in the context of MBT is time behavior with 66 instances\footnote{We mean by instance how many times it appeared per category rather than per paper} in terms of both testing level and model used (Figure~\ref{fig:sms performance testing model}).
Resource utilization, capacity, and speed/throughput were in close range with a median value of 10 instances in terms of both testing level and model used. Efficiency was the least studied performance aspect in the context of MBT, where it only appeared in one instance in terms of testing level and one in terms of the model used. 

We observe a similar trend analysing the requirements specifications. Time-behavior was the most common performance aspect (Figure~\ref{fig:srs performance application}). Out of the 149 extracted performance requirements, time behavior was specified in (71) requirements (e.g., \emph{The system shall be able to search for a specified product in less than 1 second.}\footnote{0000-gamma (the id of the SRS)
}), followed by capacity (38) (e.g., \emph{The system must handle at least 100 concurrent users and their operations}\footnote{2008-fiber}), speed/throughput (18) (e.g., \emph{The system shall be able to retrieve 200 products per second.}\footnote{0000-gamma}), efficiency (13) (e.g., \emph{Management – all management software functions shall take optimal advantage of all language, compiler and system features and resources to reduce overheads to the minimum practical level.}\footnote{2002-evla back}) and resource utilization (9) (e.g., \emph{The FTSS software and the VxWorks operating system, together shall [SRS193] utilize no more than 3 megabytes of ROM.}\footnote{2000-nasa}). 

We can see from Figure~\ref{fig:srs performance application} that most of the SRS documents with performance requirements were written for web applications, followed by real-time and embedded systems. There was a diversity in terms of performance aspects in the specified requirements for web applications, whereas for real-time systems and embedded systems the specified requirements were mostly related to time behavior. A similar observation can be made by looking at Figure~\ref{fig:sms performance aspect and application} where web application and embedded system appeared in 22 instances each and real-time systems in 19 instances. The importance of performance in web application, embedded systems and real-time system is not surprising. In a web application a large number of application users are distributed and use different communication media to access the application. Embedded and real-time systems are crucial to perform optimally, since a safety hazard could arise if performance is not addressed. For instance, in self-driving cars the time behavior for reading the value of a sensor is crucial and needs to be specified explicitly, allowing the product to be tested against that specification.

In both the identified primary studies and the reviewed SRSs, time behavior was the most common performance aspect. Nonetheless, the other performance aspects are also relevant, since they appeared in a median of 10 instances each (except efficiency) and specified in 78 requirements combined. That said, we should consider all the performance aspects when modeling performance requirements. Efficiency was the least studied (found in one paper~\cite{johnsen_experience_2017}), and the least quantified in (3) requirements (Figure~\ref{fig:srs performance with all reqs and testable}). However, efficiency was specified in (13) requirements, from which we conclude that efficiency is difficult to document and quantify. We found few examples of quantified efficiency requirements: 1) \textit{The external server data store containing RLCS status for use by external systems shall be updated once per minute}\footnote{2004-rlcs}, and 2) \textit{The system must accomplish 90\% for transactions in less than 1 second.}\footnote{2008-viper}. The examples show that it is possible to quantify efficiency. In the first requirement "only once every minute" and in the second "90\%...less than 1 second". Both combine two performance aspects, i.e., capacity and time behavior.

Looking at testing levels, performance testing research seems to focus on system level testing (Figure~\ref{fig:sms performance testing model}). This observation coincides with the notion that software performance is not associated with a single function, but rather associated with the overall system and influenced by its structure. This is also shown in the performance requirements models used in MBT. The purpose of those models is to verify the overall system behavior, e.g., timed-automata~\cite{maalej_automated_2013,maalej_model-based_2012,schumi2017checking} and behavior models~\cite{abdelgawad_model-based_2017,al-tekreeti_test_2018}. 

We extracted 50 performance requirements models and categorized them into 11 main categories (Figure~\ref{fig:sms performance testing model})~\footnote{The clustering of those 50 models into 11 clusters is available online~\cite{additionalData}.}. All 11 categories had models which were used to model time behavior requirements. The purpose of those models is to verify if the written requirements are met. This is accomplished by comparing the testing results with the corresponding performance requirements.

The most studied models were timed-automata and UML-related diagrams. Timed-automata were used to model and analyze the time behavior by measuring time differences between different states, which can model and verify time behavior aspects of software performance. However, timed-automata models have two main drawbacks. First, the models do not make the factors influencing performance explicit, which is needed to generate better test cases for performance requirements. Second, timed-automata can only model time behavior and are unable to cover other performance aspects~\cite{garousi_fault-driven_2011}, and are therefore only adequate when time behavior is the only performance aspect that needs to be tested. As we can see from our analysis of SRSs, time behavior is seldom the only performance requirement. UML-based models use an annotation approach to make the performance requirements more intuitive and the system behavior more understandable~\cite{abdelgawad_model-based_2017}. UML-based models solely document performance requirements, and are not used for test case generation of performance requirements. In many cases where UML is used, the performance requirements (e.g., time behavior, or capacity) is set on the model as annotation, which is later used during the test generation to add an extra assert to check this requirement. This model annotation is beneficial to verify the performance constraints of a functional requirement in a test-environment (machine resources and test data).

The models and frameworks that we extracted during the SMS were mostly newly developed with little to no validation~\cite{gambi_iterative_2013,gangadharan_fast_2009,loding_timed_2010,wilke_vision_2011} as seen in Figure~\ref{fig:sms type method contribution} (a). Although 31 case studies exist that validate those models (e.g., timed-automata), researchers still develop new performance requirements models and testing frameworks (Figure~\ref{fig:sms type method contribution} (b)). The reasons for developing those models and frameworks are various:

\begin{enumerate}
    \item Model-based performance testing in a specific field has not been done before, e.g., robotics~\cite{abdelgawad_model-based_2017}, self-adaptive systems~\cite{weyns_towards_2012} and cloud API~\cite{wang_model-based_2017}, has not been studied for a specific performance aspect, e.g., resource utilization~\cite{al-tekreeti_test_2018,iyenghar2011applicability,wilke_vision_2011} or time behavior~\cite{luthmann_modeling_2017}, or has not been proposed in a particular development stage, e.g., early before a prototype is created~\cite{li_quality_2018}, or late during run-time~\cite{saadatmand_testing_2013}.
    
    \item Issues associated with human factors where it is difficult to understand the model~\cite{bernardino_canopus_2016,chimisliu_abstracting_2011}, it takes extra effort to create the model~\cite{abbors_model-based_2013,siegl_partitioning_2015}, or the current approaches are prone to human error~\cite{rodrigues_pletsperf_2015}.
    
    \item The lack of automation in the current MBT approaches~\cite{enoiu_model-based_2013,maalej_conformance_2012,vain_multi-fragment_2017,wang_system_2017}
    
    \item Others reasons, e.g., using petri nets to model time behavior aspects~\cite{camilli_event-based_2017}.
\end{enumerate}

A majority of the analyzed papers (46) suggest a new concept or framework for MBT, using formal theory research (Figure~\ref{fig:sms type method contribution}). This set is followed by 41 papers conducting field studies and field experiments that aim at validating the new model presented in the same paper. This focus on theoretical work and studies in a relative controlled environment is another indication that the models are not validated under realistic conditions, as also observed by Prenninger et al. in their review of eight case studies on MBT~\cite{hutchison_15_2005}. A similar observation can be made by looking at the contribution of theses papers in Figure~\ref{fig:sms type method contribution} (b) where most papers introduced new ideas and methods rather than evaluating pre-existing models. It would be crucial to evaluate those models, as the lack of evaluation of MBT techniques poses a risk factor of using those techniques in industry practice. This factor influences the techniques' reliability, and evaluated techniques would positively affect their adoption in future software projects~\cite{dias-neto_picture_2010}.

\subsection{Implications of the SMS on performance requirements in MBT}
We gained useful insights on performance requirements modeling in the context of MBT by conducting the SMS. First, performance requirements that were not studied before, (e.g., resources and speed/throughput), gained interest in recent years, as seen in Figure~\ref{fig:sms studies per year}. This is an indication that more research is required in these aspects. Second, some performance attributes (e.g., time behavior) were used as test verdicts~\cite{rodrigues_pletsperf_2015}, while others (e.g., capacity) were used as a foundation to the test environments~\cite{iqbal_environment_2015}. Third, performance requirements could be modeled separately from functional requirements, and test environments could be generated from the model~\cite{abbors_approaching_2010}.

However, we argue that the performance requirements models found by our SMS (Figure~\ref{fig:sms performance testing model}), do not satisfy all goals of MBT simultaneously, i.e., support requirements validation, ensure requirements testability, and support test case generation. Therefore, we developed PRO-TEST to aid the model-based performance testing process, which we introduce next.

\section{PRO-TEST}\label{sec:pro-test}
\label{chap:performance relational model}
In this section, we introduce and evaluate the Performance Requirements VerificatiOn and Test EnvironentS generaTion approach (PRO-TEST). PRO-TEST aims at checking the completeness and correctness of performance requirements and at generating the parameters of test environments. 

\begin{figure}[htb]
\centering
    \includegraphics[width=0.9\columnwidth]{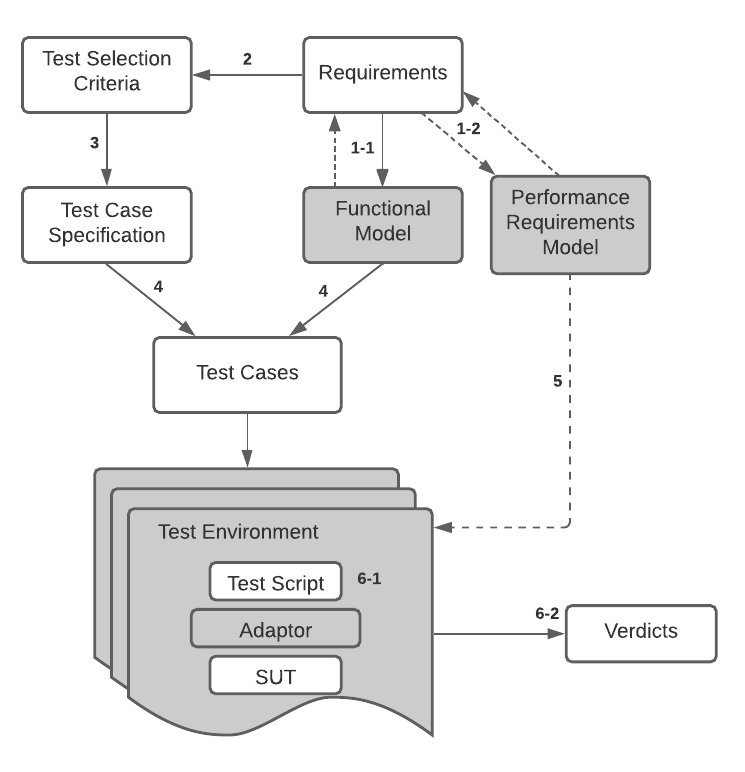}
\caption{MBT Process in the context of performance testing}
\label{fig:mbt process performance}
\end{figure}

Figure~\ref{fig:mbt process performance} illustrates the MBT process in the context of performance testing. The figure is a modified version of Utting et al.'s~\cite{utting_taxonomy_2012} MBT process diagram that we introduced in Section~\ref{sec:MBT}. The modified process steps are shown with dotted arrows, and the modified/added artifacts are filled in grey color. We made three modifications to the diagram. First, we split the step of requirements modeling into two sub-steps: functional modeling (1-1) where a model is created from the functional requirements, and performance requirements modeling (1-2) where performance requirements are modeled. Second, we added an iterative process between the requirements and the created models (functional and performance). This change underlines how MBT supports requirements validation (an MBT goal). The modeling stage should detect requirements issues and changes should be made to the requirements to fix these issues. Third, we added a new \textit{Step 5}, in which test environments are generated from the performance requirements model. The software performance is thereby directly related to the test environment. Setting up a test environment requires specifying setup parameters (e.g., capacity of users) and metrics parameters (e.g., response time). These parameters are derived from the performance requirements models. 

In summary, PRO-TEST consists of 1) performance requirements modeling and 2) test environment generation. These activities correspond respectively to step 1-2 and step 5 in Figure~\ref{fig:mbt process performance}. The approach is not meant to be used as standalone but rather accompanied by any MBT approach that generates functional test cases, which results in functional test cases mapped to test environments that test the performance of the SUT. 

We illustrate PRO-TEST's core concepts in Section~\ref{sec:pro-test-description} and explain the steps and guidelines for creating the performance requirements model in Section~\ref{sec:creating-PRO-TEST}. Additionally, we explain the steps of generating test environments in Section~\ref{sec:generating test environments}, illustrate its application on an example in Section~\ref{sec:example-PRO-TEST}, and apply PRO-TEST on a set of 149 performance requirements in Section~\ref{sec:ss-model-eval}, discussing the strengths and weaknesses of the approach.

\subsection{PRO-TEST approach development and description}
\label{sec:pro-test-description}

We intend to propose a modeling approach that addresses the limitations of current model-based performance testing. Specifically, the modeling of the five performance aspects in Section~\ref{sec:sw-performance} to verify the requirements while generating test environments. Looking at the existing approaches that we identified in our SMS, we found that performance requirements affect test environments. Therefore, instead of creating an approach that models both performance and functional requirements, we chose to develop an approach that focuses on modeling performance requirements and generating test environments. This approach can be accompanied by existing well-established MBT approaches that already handle functional modeling and testing. By focusing on performance requirements, we increase the chance of our approach being used by practitioners who are already using existing MBT for functional testing, without the need to replace their existing tools but add to what they already use.

The development of PRO-TEST was inspired by two related principles. First, the experiment principle that illustrates the relationship between dependent and independent variables~\cite{wohlin_experimentation_2012}. Second, cause-effect graphs (CEGs)~\cite{elmendorf1973cause} that can be used to model the relationships between causes and effects.

\begin{figure}[htb]
\centering
    \includegraphics[width=0.9\columnwidth]{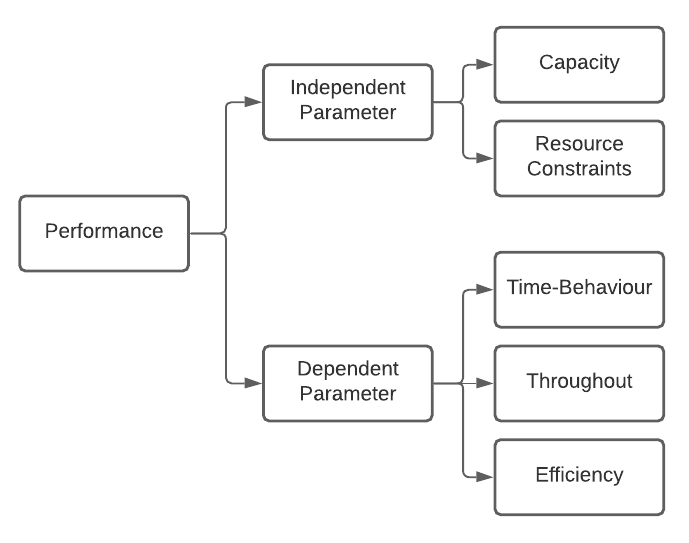}
\caption{Performance Parameters Taxonomy}
\label{fig:Performance Parameters Taxonomy}
\end{figure}

We analyzed the different performance aspects while having the cause-effect concept in mind. The main insight we had is that one set of performance requirements (capacity, resource constraints) can influence another set (time behavior, throughput, efficiency). This concept is shown in a taxonomy tree (Figure~\ref{fig:Performance Parameters Taxonomy}) that classifies the aspects in independent and dependent performance parameters. The independent parameters consist of capacity (e.g., the maximum number of users), and resource constraint (e.g., storage size), which represent constraints on the software. The dependent parameters consist of time behavior (e.g., response time), throughput/speed (e.g., requests per time unit), and efficiency (e.g., response time in regards to memory size), and are measurements of the software performance. The manipulation of the independent parameters causes changes in the dependent parameters. For example, if we require the system to use fewer resources (all other things being equal), it would lead to a higher response time, lower throughput, or efficiency. The purpose of this taxonomy tree is to identify which performance requirements are the influencing factors and which ones are impacted, as this is important to distinguish when modeling testable system requirements. The taxonomy tree is by no means exhaustive, but rather a classification of the most common (studied and specified) performance requirements.

In the previous paragraph, we used the term resource "constraints" instead of "utilization" in order to emphasize the interpretation of the parameters as an independent parameter. Looking at our results from the SRS analysis, we found that the specified resource utilization requirements could be both a dependent variable that we measure when we run the tests or an independent variable that affects the dependent variables when constructing and running the tests. For example, if we take the requirement \emph{"The FTSS software and the VxWorks operating system, together shall [SRS193] utilize no more than 3 megabytes of ROM."}~\footnote{2000-nasa}, there are two methods to test it. Firstly, we run the tests, measure the utilized ROM, and make sure the software does not utilize more than 3 megabytes. Alternatively, we set up a test environment with 3 megabytes of ROM  as a constraint, run the tests, and if the tests run completely, then the software satisfies this requirement. We chose to apply the second method (hence the use of terminology resource constraints) since it works better when the specified requirement affects our decision when setting up the test environment. For instance, to test the requirement \emph{"GParted is not a resource hog and will run on almost every computer"}~\footnote{2010-gparted}, we can't run the tests and measure the utilized resources (even if this requirement is to be quantified). Instead we need to define a set of representative computers and run the tests on them.

\begin{figure}[htb]
\centering
    \includegraphics[width=0.9\columnwidth]{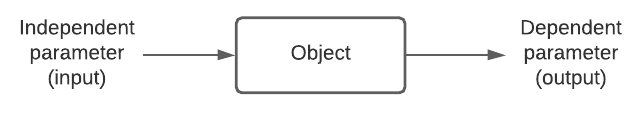}
\caption{Performance requirements model}
\label{fig:PRO-TEST-model}
\end{figure}

Figure~\ref{fig:PRO-TEST-model} presents the main components of a performance requirements model. The model consists of three main parts: 
\begin{enumerate}
    \item The object element referring to the SUT or part of it, i.e., a function that has the performance requirements associated with it.
    \item The independent parameters which act as inputs. They affect the test environment where the test runs and affect the test data.
    \item The dependent parameters which act as outputs. They are the metrics or results of running the tests, used to compare the test results with the written performance requirements.  
\end{enumerate}

Performance requirements are modeled with PRO-TEST using the taxonomy tree that acts as a guide when extracting, categorizing, and finding  missing performance requirements.

\subsection{Performance requirements model}
\label{sec:creating-PRO-TEST}

There are three steps that should be followed when modeling the performance requirements of the software. 

\begin{itemize}
    \item \textit{Step 1: Define the objects.} Look up the object that the performance requirements on hand applies to. The objects could be the system, specific functions, or a collection of functions.
    \item \textit{Step 2: Define the independent and dependent parameters.} Extract the performance parameters from the requirements, and code them with the appropriate performance aspect using the taxonomy tree. Then add those parameters to the corresponding model. 
    \item \textit{Step 3: Compare the model with the taxonomy tree.} Take the created performance requirements model and compare it with the taxonomy tree. Look for any possible missing parameters. If some parameters are missing, look for the possibility of merging models with the same object. If there are still some missing parameters, then there is a problem with the requirements. Check with requirements engineers or customers to negotiate the requirements. Otherwise, the model is complete and the specified requirements are quantified and can be tested. When the modeling is done, the next step is to design the test suite.
\end{itemize}{}

When using PRO-TEST with performance requirements, one should take into consideration the following guidelines which help to model the requirements.

\begin{itemize}
    \item \textit{Guideline 1: Verify the completeness of the requirements.} Check the relation between different requirements. There should be a correspondent independent input for each dependent output. Having one without the other would result in ambiguous requirements, which would reflect an incomplete performance requirements model.
    \item \textit{Guideline 2: Verify feasibility.} The requirement should fit with one of the performance aspects' definitions in Section~\ref{sec:sw-performance}.
    \item \textit{Guideline 3: Verify quantifiability.} Each requirement should have a quantity that describes the target level of performance, and an object that specifies where the target level applies (system, a specific function, or a collection of functions).
    \item \textit{Guideline 4: Specific condition.} Check if the requirements apply in specific circumstances or scenarios. The performance requirements might have the same objects but under different conditions, i.e., peak time. In this case, one should make a different model for each of those conditions, because each condition has different parameters that apply to the test environment and different measurement levels.
    \item \textit{Guideline 5: Mandatory performance aspects.}
    To generate meaningful test environments, each model requires the following performance aspects to be specified: 1) capacity and resource constraints to help set up the test environment, and 2) time behavior or throughput which acts as the metric to measure when running a test.
\end{itemize}{}

While these suggestions stem from our experience of modeling nearly 150 performance requirements from 34 SRS documents, they are not exhaustive and should not be considered as rules. 

\subsection{Generating Test Environments}
\label{sec:generating test environments}

One of the goals of PRO-TEST is to generate test environments, which aids the verification of performance requirements in the SUT. As seen in Figure~\ref{fig:mbt process performance}, the generated test environments are required to run performance test and affect the outcome of performance tests.

Using the created performance requirements models, we generate parameters for the test environments. These parameters are divided into two groups: \emph{constraints} and \emph{metrics}. The \emph{constraints} parameters are required to set up the test environments and stem from the independent parameters in the taxonomy in Figure~\ref{fig:Performance Parameters Taxonomy}. The \emph{metrics} parameters are indicators for the success or failure of the test cases run in the test environment, and stem from the dependent parameters in the taxonomy in Figure~\ref{fig:Performance Parameters Taxonomy}.

\begin{lstlisting}[caption=Test Environment Generation Algorithm, label={lst:test environment generation algorithm}]
Create constraintsList
Create metricsList
Add resource constraints to constraintsList
Add capacity to constraintsList
Add time behavior to metricsList
Add speed/throughput to metricsList
Add efficiency to metricsList

Create environmentsList
CALL testEnvGenerator with constraintsList and metricsList
Add the generated environment to the environmentsList
FOR each constraint in constraintsList
    CALL testEnvGenerator with constraint and metricsList
    Add the generated environment to environmentsList
END FOR

CALL mapTestCasesToEnvironments with environmentsList
\end{lstlisting}

We show in Listing~\ref{lst:test environment generation algorithm} the algorithm to generate the test environments that will be used to run the test cases. The algorithm consists of three main steps. 1) Create two lists of parameters, \emph{constraintsList} and \emph{metricsList}, and add the parameters from the created performance requirements models to the corresponding list based on the classification in the taxonomy tree. 2) Create an \emph{environmentsList}, one for each parameter in the \emph{constraintsList} with all parameters in the \emph{metricsList}, and an environment where all parameters in \emph{constraintsList} and \emph{metricsList} are included. 3) Map the test cases to the created environments in \emph{environmentsList}. The test cases mapped to the test environments are those that verify the object (e.g., function) to which the performance requirements refer.

To automatically generate test environments from the created performance requirements model, we implemented a Python script \footnote{The test generation script is available online~\cite{additionalData}}. The script takes as input the list of performance requirements models (in CSV format) created by the tester. The output of the script is a list of test environments (in JSON format). Each test environment consists of a list of constraints to construct the test environment and object-metric pairs that indicate what functions should be tested and measured in this environment. We chose JSON as output format since it is a widely used in practice. Generating test environments in this format makes it fairly easy to adapt to different testing tools.

\subsection{Example of PRO-TEST}
\label{sec:example-PRO-TEST}
To illustrate PRO-TEST, we present an example, following the three steps described in Section~\ref{sec:creating-PRO-TEST} for creating the performance requirements model. Then, we generate parameters for test environments following the test environment generation presented in Section~\ref{sec:generating test environments}. We extracted performance requirements for a telescope control software shown in Table~\ref{tab:demonstration requirements}.

\begin{table*}[htb]
\caption{Example Performance Requirements for PRO-TEST Approach Demonstration}
\label{tab:demonstration requirements}
\centering
    \begin{tabular}{p{0.05\textwidth}p{0.60\textwidth}p{0.25\textwidth}}
      \hline\noalign{\smallskip}
        \textbf{No.} & \textbf{Performance Requirements}  & \textbf{Performance Aspect} 
        \\
        \noalign{\smallskip}\hline\noalign{\smallskip}
        PR1 & The Gemini \underline{software} should have no hard restrictions on the number of \textbf{simultaneous users}, but should allow for policy decisions that do restrict the amount of simultaneous access. & Capacity
        \\
        PR2 & Every \underline{command must be accepted/rejected} within \textbf{2 sec} and before the corresponding action occurs. (This is different than the ACK/NAK response of the communications protocol - here, the target system must have examined the command and verified its validity.) & Time Behavior
        \\
        PR3 & \underline{Status display update} must be within \textbf{4 sec} at the local stations (certain functions, such as telescope position, may have tighter constraints). Remote station update response is given in the Requirements for Remote Operations section. & Time Behavior
        \\
        PR4 & \underline{Requests of subsystems for status information} must be answered within \textbf{5 sec} and be possible in maintenance level operation.  & Time Behavior
        \\
        PR5 & Requirements for response times within the \underline{user interfaces} are given in the User Interface requirements section. & Time Behavior
        \\
        PR6 & The \underline{user interface} should rather be seen as a package to be callable from a \textbf{large number of stations}, depending on where a user is. & Capacity
        \\
        PR7 & The \underline{user interface} should also be \textbf{network} transparent so that it does not matter where it is being run. & Resource Constraints
        \\
        PR8 & As a conclusion, the Gemini 8m Telescopes control \underline{software} shall allow simultaneous operation of up to \textbf{six active control} nodes and up to \textbf{two more monitoring nodes} (one local and one remote) without appreciable degradation of performance. & Capacity
        \\
        PR9 & In practice the operation and facilities foreseen so far for the Gemini 8m Telescopes will limit this number to a maximum in the order of three active nodes, but the Gemini 8m Telescopes computers and \underline{software} shall be capable of coping with the load of \textbf{10 active nodes}, should the case arise. & Capacity
        \\
        PR10 & All software bugs should be logged and then fixed as soon as possible after detection. The goal is to have restart conditions occur only on hardware failure. Fault recovery, exception handling, fail-safe checks, etc. should be used to improve reliability. & Availability
        \\
        \noalign{\smallskip}\hline
    \end{tabular}{}
     \vspace{1ex}

     {\raggedright The requirements in this table were extracted from the SRS document 1995-gemini \par}
\end{table*}{}

\begin{figure}[htb]
\centering
    \includegraphics[width=0.7\columnwidth]{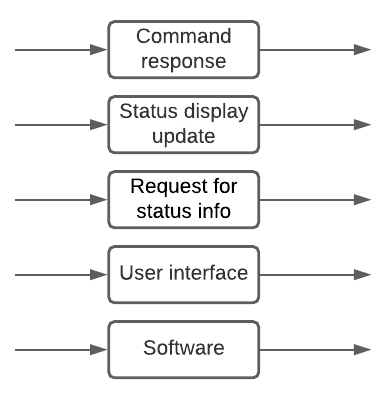}
\caption{PRO-TEST Example --- Step 1}
\label{fig:PRO-TEST example step1}
\end{figure}

\subsubsection{Performance requirements model}

\paragraph{Step 1: Define the objects.} We defined five objects from the requirements: command response, status display update, request for status info, user interface and software. Then we created five models, one for each object as shown in Figure~\ref{fig:PRO-TEST example step1}.

\begin{figure}[htb]
\centering
    \includegraphics[width=0.9\columnwidth]{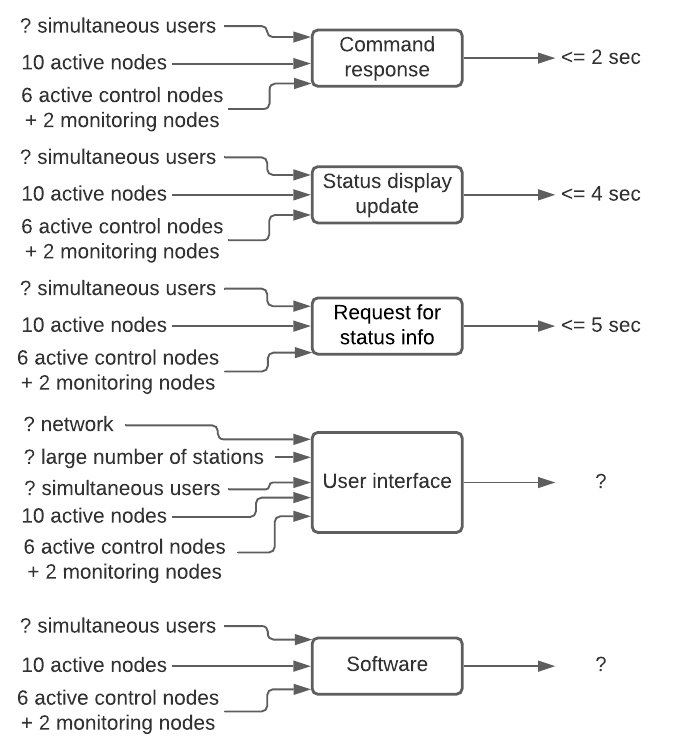}
\caption{PRO-TEST Example --- Step 2}
\label{fig:PRO-TEST example step2}
\end{figure}

\paragraph{Step 2: Define the independent and dependent parameters.} We extracted the performance parameters (10 active nodes, large number of stations, simultaneous users, 6 active control nodes \& 2 monitoring nodes, $\leq$ 2 sec, $\leq$ 4 sec, $\leq$ 5 sec and network) from the requirements, and coded them with the related performance aspects as per the taxonomy tree. We present the associated performance aspect in the last column of Table~\ref{tab:demonstration requirements}. Then we added those parameters as independent and dependent parameters in the model as shown in Figure~\ref{fig:PRO-TEST example step2}. At this stage we identified four issues in the requirements:
\begin{enumerate}
    \item PR10 is an availability requirement, which is not to be found in our taxonomy tree (guideline 2), hence we exclude PR10. 2) PR5 indicates that there should be a time behavior requirement for the user interface. However, we examined the SRS document and we did not find any time behavior requirements for the user interface. Hence, PR5 can not be modeled and it indicates a missing requirement. 
    \item PR1 (simultaneous users), PR6 (large number of stations), and PR7 (network) are not quantified (guideline 3). 
    \item PR6 is ambiguous as "without appreciable degradation of performance" is not unclear. 
    \item PR8 and PR9 are conflicting requirements. PR8 specifies a capacity of 8 nodes (6 active plus 2 monitoring), however, PR9 specifies a capacity of 10 active nodes. 
\end{enumerate}

\begin{figure}[htb]
\centering
    \includegraphics[width=0.9\columnwidth]{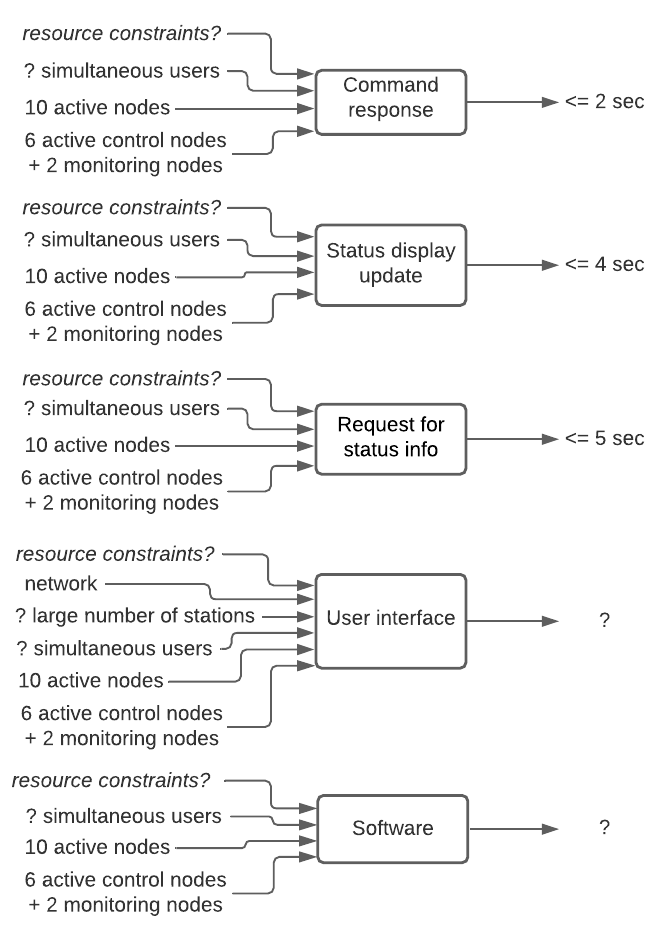}
\caption{PRO-TEST Example --- Step 3}
\label{fig:PRO-TEST example Step3}
\end{figure}

\paragraph{Step 3: Compare the model with the taxonomy tree.} We compared the created model with the taxonomy tree to identify any possible missing parameters. We put the possible missing requirements on each corresponding model as seen in Figure~\ref{fig:PRO-TEST example Step3}. Resource constraints parameters are missing from the models and the specified requirements for the software since there were no requirements indicating resource constraints. Another issue is that the requirement PR5 (large number of stations) applies to other parts of the system as well (missing requirement). Moreover, there are no performance requirements from the dependent parameters (time behavior, speed/throughput, or efficiency) that apply to the software or the user interface. 

At this point of the analysis, the identified issues should be discussed with the requirements engineers or customers to negotiate the requirements and fix the issues: asking for 1) the missing requirements, 2) quantify PR1, PR6, and PR7, 3) clarify or reformulate the existing requirement PR8 into two requirements, one that specifies the capacity for the software, and the other that specifies the dependent parameter e.g., time behavior, and 4) resolve the conflict in the requirements PR8 and PR9.

\subsubsection{Test environments generation} 
We generate test environments parameters following the test environment generation algorithm presented in Listing~\ref{lst:test environment generation algorithm}. We feed each model in Figure~\ref{fig:PRO-TEST example Step3} (constraints and metrics) to the algorithm as input, and as output we get a set of environments (one per constraint and one with all constraints). This makes debugging easier, as the tester can identify the troublesome performance constraint(s) just by looking at the constraint(s) used to construct the test environments in which the failed test was run. 

\begin{figure}
    \centering
\scalebox{0.6}{\begin{forest}
  for tree={
    font=\ttfamily,
    grow'=0,
    child anchor=west,
    parent anchor=south,
    anchor=west,
    calign=first,
    edge path={
      \noexpand\path [draw, \forestoption{edge}]
      (!u.south west) +(7.5pt,0) |- node[fill,inner sep=1.25pt] {} (.child anchor)\forestoption{edge label};
    },
    before typesetting nodes={
      if n=1
        {insert before={[,phantom]}}
        {}
    },
    fit=band,
    before computing xy={l=15pt},
  }
[test environment
    [constraints 
        [0 [description] [att class]]
        [1 [description] [att class]]
        [...]
    ]
    [object metric pairs
        [0 [object]
        [metric 
            [description] 
            [att class]]
        ]
        [1 [object]
        [metric 
            [description] 
            [att class]
        ]]
        [...]
    ]
    [errors]
]
\end{forest}}
    \caption{Test environment JSON file structure}
    \label{fig:test environment json file structure}
\end{figure}
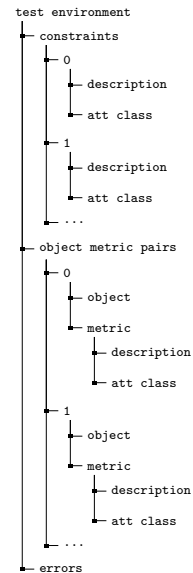

We used our test environment generations script to automatically generate test environments from the created models. In Figure~\ref{fig:test environment json file structure}, we show the structure of the generated file. The root node of the file contains an array of generated test environments. Each test environment consists of a list of \emph{constraints} and a list of \emph{object-metric pairs}. A \emph{constraint} presented using a \emph{description} and an \emph{att class} (the performance aspect). An \emph{object-metric pair} consists of the object to be tested (e.g., a function), and the metric to be measured. A metric is presented using a \emph{description} and \emph{att class}. Errors in the modeled performance requirements will be shown in \emph{errors}.

The results of generating test environments can be found in Table~\ref{tab:PRO-TEST example test environments}. The rows 1-4 can be used to construct test environments. This is not possible for the remaining rows (5-8), as they are missing constraints and/or metrics. For instance, the question mark in row 5 for the constraint \emph{simultaneous users} is an indication of a missing quantity of \emph{simultaneous users}. As we mentioned earlier in this section the requirements should be negotiated with the customer, so we can fill the gaps in our tests.

\begin{table*}[ht]
\caption{PRO-TEST Example - Test Environments Summary}
\label{tab:PRO-TEST example test environments}
\centering
    \begin{tabular}{lp{0.30\textwidth}p{0.60\textwidth}}
        \hline\noalign{\smallskip}
        Id & Constraints & Object (Measure)
        \\
        \noalign{\smallskip}\hline\noalign{\smallskip}
        1 & 10 nodes & Command Response ($\leq$ 2 sec), status display update ($\leq$ 4 sec), request for status info ($\leq$ 5 sec), software
        \\
        2 & 100 simultaneous users & Command Response ($\leq$ 2 sec), status display update ($\leq$ 4 sec), request for status info ($\leq$ 5 sec), software
        \\
        3 & 10 nodes, 100 simultaneous users & Command Response ($\leq$ 2 sec), status display update ($\leq$ 4 sec), request for status info ($\leq$ 5 sec), software
        \\
        4 & 10 nodes & User interface
        \\
        5 & ? simultaneous users & User interface
        \\
        6 & ? network & User interface
        \\
        7 & ? number of stations & User interface
        \\
        8 & 10 nodes, ? simultaneous users, ? network, ? number of stations & User interface
        \\
        \noalign{\smallskip}\hline
    \end{tabular}{}
\end{table*}{}

\subsection{Sample study - model evaluation}
\label{sec:ss-model-eval}

We applied PRO-TEST on 34 SRS documents from the SRS collection. We extracted in total 149 performance requirements from the SRS documents, i.e., requirements that fit the definition of performance aspects in Section~\ref{sec:sw-performance}.

We extracted the performance requirements from the SRS collection and applied PRO-TEST by modeling the requirements as explained in Section~\ref{sec:creating-PRO-TEST}. We did not generate test environments from the created performance relational models, since test environments generation would be more meaningful if used with another MBT approach to generate test cases from functional requirements. This is outside the scope of this paper.

In Table~\ref{tab:PRO-TEST-evaluation-results} we present two types of defects found by PRO-TEST. The first defect is related to quantifiability. We found that 106 out of 149 requirements were quantified, while the remaining 43 were quantifiable but were not actually quantified (e.g., \textit{"The product will reside on the Internet so more than one user can access the product and download its content for use on their computer."}\footnote{2001-space fraction}). 

\begin{table}[ht]
\caption{PRO-TEST evaluation results}
\label{tab:PRO-TEST-evaluation-results}
\centering
    \begin{tabular}{ll}
        \hline\noalign{\smallskip}
        Defect & Quantity
        \\
        \noalign{\smallskip}\hline\noalign{\smallskip}
        Not-quantified Requirements & 43
        \\  
        Under-specified Parameters & 180
        \\
        Under-specified Resource constraints & 100
        \\
        Under-specified Capacity & 39
        \\
        Under-specified Time-behavior & 22
        \\
        Under-specified throughput & 19
        \\
        Under-specified Efficiency & 0
        \\
        \noalign{\smallskip}\hline
    \end{tabular}{}
\end{table}{}

The second type of defect is related to under-specified or missing requirements. We found a total of 180 missing parameters in the analyzed requirements. The majority of them (100) were related to resource constraints, followed by capacity (39), time behavior (22), and throughput/speed (19). No missing parameters for efficiency requirements were detected. As defined in Section~\ref{sec:sw-performance}, efficiency is a combination of more than one parameter. Hence, to some extent, the existence of those parameters (e.g., time behavior and resources constraints) eliminates the need for efficiency requirements.

In the included SRS documents there were 204 performance requirements, categorized by the original author of the SRS; We identified and categorized 132 of those requirements, while we could not fit 67 requirements to any of the performance aspects definitions in Section~\ref{sec:sw-performance}. For example, \textit{"Assuming submitted statistics for jobs are accurate, the Libra scheduler will ensure that all jobs are completed with a 10\% error allowance."}\footnote{2001-libra}. Other requirements were hard to understand how they fit in performance requirements, e.g., \textit{"The database retrieval and update response time shall not impact any other performance requirements such as the GUI response time or monitoring and control responses."}\footnote{2004-rlcs}. This requirement mentions response time, but it does not clearly state where does it apply or what the target level of performance is. There were some requirements that were more difficult to identify, e.g., \textit{"The HATS-GUI shall allow a user to request transformations while HATS-SML is performing transformations or parsing."}~\footnote{2001-hats}. It could be argued that this requirement is an efficiency requirement. But reading it carefully we concluded that this is not a performance requirement, but rather a usability requirement that demands parallel processing or multitasking. According to Ho et al.~\cite{ho_agile_2006}, a performance requirement can be categorized into four levels (0 to 3). These levels show the maturity, suitability, and validation of performance requirements. Based on their definition, this requirement is classified as level 0 (lowest), which is descriptive and can only be evaluated qualitatively. The requirements in this paragraph were extracted from 2001-libra, an SRS for economy-driven cluster scheduler for high-performance clusters, 2004-rlcs, an SRS for an interstate reversible lane control system, and 2001-hats, a high assurance transformation system. Relying solely on a qualitative evaluation of performance in these systems leads potentially to unsatisfied customers. 

Out of the 149 requirements, 43 were not quantified. Those requirements fall into two categories. (1) Requirements with minor issues, i.e., just missing the numerical value. For example \textit{"The tools shall be able to scale to process large collections using distributed processing and data transport."}\footnote{2009- warc III}. This is a capacity requirement, that applies to the whole tool (object). However, the size of the collection is not defined; it could be 100 or 100,000. Since the requirement does not specify a range, we do not know how to test it. (2) Requirements with major issues. For example \textit{"Loading speed: The data system shall load as quickly as comparable productivity tools on whatever environment it is running in."}\footnote{2006-stewards}. This requirement refers to efficiency in an ambiguous manner: "as quick as possible" and "on whatever environment". No test could be written to verify if the system satisfies this requirement.

Performance aspects were not considered equally by the requirements engineers when writing the SRS documents, which shows the lack of knowledge in the inter-dependency relation between different aspects as shown by PRO-TEST. 100 out of 109 created models had missing requirements in resource constraints. It could be argued that resource constraints are not a part of performance requirements. However, it does affect software performance, and there were some SRS documents that specified resource constraints properly e.g., \textit{"The Framework Shell SHOULD NOT utilize more than 40 megabytes of RAM."}\footnote{2005-znix}. 

We generated 96 test environments from the performance requirements models that we created from the SRS collection. All of the generated test environments had missing or unquantified requirements.

Bondi~\cite{bondi_best_2012} suggested that a performance requirement should have nine characteristics: unambiguous, measurable, verifiable, complete, correct, mathematically consistent, testable, traceable, and can be linked to business and engineering needs. Our study corroborates that PRO-TEST supports a subset of these characteristics: it helps engineers in verifying performance characteristics as it makes lack of information explicit (completeness and quantifiability), it detects unclear information (ambiguity), and associates performance requirements to test environments.

\section{Discussion} \label{chap:pro-test discuss}
In this section, we discuss the different aspects of PRO-TEST. We compare the performance taxonomy and the performance aspect inter-dependency relation with those from the literature, list the limitations of the approach, discuss how the approach differs from other MBT approaches, show our observations regarding performance requirements, and finally we discuss PRO-TEST with performance prediction.

\subsection{Previous performance aspect classifications}

As we saw from our SMS and SRS analysis results, five performance aspects were studied and used in practice. Thus, testers should consider these aspects when testing software performance. Eckhardt et al.~\cite{eckhardt_challenging_2016} specify a template to write performance requirements. They considered three aspects of performance requirements, namely time behavior, throughput and capacity. In addition, they specified performance context (e.g., platform, measurement location and load) as part of each requirement. However, they do not consider resource constraints, but rather the platform (hardware) under which the requirement applies. It is seldom the case that specifying hardware requirements is enough to test system performance and ensure the desired time behavior, throughput and efficiency. For instance, smartphone applications, vehicles software, cloud services, and desktop applications, all share resources with other applications running on the same platform. In this case, performance testing verdicts are more reliable when we specify the available resources for the system rather than the platform it runs on. Nixon et al.~\cite{nixon_management_2000} categorized performance requirements into time (response time, throughput and management time) and space (main memory, secondary storage). They did not account for capacity which we consider in our taxonomy tree. 

\subsection{Performance aspects inter-dependency}

The dependency relation between the five performance aspects as far as we know was not observed before. Cai et al.~\cite{cai_addressing_2002} considered two aspects of performance, time and space, and called the relation between these aspects side-effects. They did not define clearly the nature of the effect, nor considered the other performance aspects. Eckhardt et al.~\cite{eckhardt_challenging_2016} proposed that each specified performance requirement should have platform and load in the same requirement, since these aspects affect all other performance aspects. They do not consider the case when platform and load requirements are specified in separate requirements, which can be the case as we saw in our SRS analysis results.

\subsection{PRO-TEST benefits}
Using PRO-TEST to model performance requirements and generate test environments has the following benefits: 
\begin{enumerate}
    \item It helps software engineers to understand the requirements better. When the performance requirements are visualized and by using the taxonomy tree, it becomes easier to find the relation between the requirements and how they relate to functional requirements
    \item It acts as a validation tool for the requirements. By modeling the performance requirements, we can find out 1) if there are issues with some requirements, which can not be modeled, and 2) if other requirements are missing.
    \item It informs software testers in what environments the tests should be run. This saves time and resources as it allows testers design efficient test suites.
\end{enumerate}

\subsection{PRO-TEST limitations}\label{sec:pro-test limitations}
There are some limitation of using this modeling approach to model performance requirements. First, the taxonomy tree is rather abstract. By using the taxonomy, we can identify that capacity requirements are missing, however, currently it provides no support or details about what is missing, e.g., data, users, requests. These could be specified in more detail in further nodes of the taxonomy. Second, the approach is prone to human error. Since the extraction and coding of the parameters is done manually, the process depends on the engineers' interpretation of the requirements. This could be avoided by automating the process using natural language processing. Fourth, a lack of inspection of the requirements' quality. As argued by Bondi~\cite{bondi_best_2012} a good performance requirement should specify to what degree a requirement should be met, i.e., we should specify if the requirement applies all the time or a specific amount of the time (99\% of the time). Using the PRO-TEST we do not detect those quality aspects of the requirements.

The main limitation of our approach of test environments generation, is that it can be difficult for a tester to debug the failed performance test. PRO-TEST generates one test environment per constraint, in addition to a test environment that aggregates all constraints. If performance tests fail in the test environment that aggregates all the constraints, then it is difficult to identify which interaction of the test constrains is the cause of the failure.

\subsection{Observations on dependent and independent parameters}
The dependent parameters (time behavior, throughput, efficiency) were more often specified than independent parameters (capacity, resource constraints) in performance requirements. This is clear from the results, where out of the 180 under-specified requirements 139 missing requirements were under the category of independent parameters (i.e., capacity and resource constraints). There could be many reasons for this outcome. First, it is possible that some requirements engineers or customers have a misconception when it comes to some performance aspects. Resource constraints could be thought of as part of hardware specifications. Second, it may be more difficult to specify those parameters during the initial stage of a software development cycle. If no prior experience exists it is difficult to asses how much resources are utilized or capacity required,  i.e., no clear estimation existed about capacity. This increases the risk of scalability issues appearing later. Similar to what happened at the PokemonGo launch~\cite{PokemonRolleOUt}, as the developers did not expect the big surge in the number of users. Third, resource constraints was left out intentionally. Today hardware virtualization is used extensively in deployed applications, it is very flexible and affordable to invest in higher specs hardware than more efficient software.

\subsection{Performance prediction}\label{sec:performance prediction}
Performance prediction is an approach to ensure the performance of the system by simulating the system behavior. Similar to MBT, performance prediction can use models to illustrate the system behavior~\cite{balsamo_model-based_2004,woodside_future_2007}. Performance prediction is used to validate the system performance early before building the system (e.g., in a simulated environment)~\cite{balsamo_model-based_2004}. In contrast, PRO-TEST verifies performance requirements through modeling and generates test environments for performance testing. 

Performance prediction is useful in systems with hardware components, where we want to understand the effect of the components used on the system performance. At the same time, the PRO-TEST and model-based performance testing approaches are appropriate to generate means of testing the software before deployment.

\section{Answering the research questions}\label{sec:arq}
We answer now our four main research questions.

\textbf{RQ1} \textit{Which aspects of performance requirements are used in MBT?}

All performance aspects presented in Section~\ref{sec:sw-performance} were used in MBT but to different extents. Time behavior was the most studied by researchers and specified by practitioners in the SRSs. Capacity, throughput, and resource constraints were studied and specified but to a lesser extent compared to time behavior. Efficiency was the least studied aspect with one paper and was only quantified in about 3 out of the 13 written efficiency requirements. We found many models that can be used to model those aspects. We can see in Figure~\ref{fig:sms performance testing model}, many of the models were used to model more than one performance aspect. 

\textbf{RQ2} \textit{How to implement MBT on performance requirements aspects?}

We found 50 models in the literature to model software performance requirements, and grouped them into 11 clusters (Figure~\ref{fig:sms performance testing model}). The purpose of those models is to document and visualize performance requirements. Those models do not satisfy the goals of MBT, which are 1) validate the specified requirements, 2) better understand those requirements, and 3) generate a suitable test suite. Hence, we developed PRO-TEST that consists of a model and a taxonomy tree for performance aspects, which verifies performance requirements and generate test environments. The performance requirements model with the taxonomy tree is not just a modeling approach for performance requirements. It is also a concept that identifies the relationship between different performance aspects.

\textbf{RQ3} \textit{To what extend is the identified approach effective at modeling performance requirements written for real-life projects?}

The results from PRO-TEST evaluation indicate that the developed approach can be used to model requirements from real-life projects. We applied PRO-TEST to performance requirements from 34 SRS documents. The approach could detect issues related to ambiguity, quantifiability and completeness of performance requirements. We could also understand the interrelation between those requirements. However, there are some limitations to PRO-TEST. 1) The taxonomy tree is not detailed enough, e.g., we do not know which type of capacity is missing (users, data size). 2) manually modeling the requirements is prone to human errors. Those limitations should be addressed to achieve the maximum benefits of MBT.

\section{Conclusions and future work}\label{sec:cfw}
In this study, we  illustrated how PRO-TEST can improve the understanding of performance requirements and support the identification of requirement defects. We conducted a systematic mapping study in the context of model-based performance testing and studied a repository of publicly available software requirements. We found from our SMS that researchers studied and modeled all performance aspects. However, there was a need to develop an approach to verify performance requirements that takes into consideration the goals of MBT. We developed PRO-TEST and showed by our evaluative study that it can be used to verify performance requirements and generate test environments. The benefits of PRO-TEST adds value to MBT. It helps software engineers to understand the requirements better, validate them, and generate test environments semi-automatically. In addition to the performance relational model, we developed the taxonomy tree, which shows the cause-effect relation between different performance aspects. 

Future work concerns more in-depth validation of PRO-TEST, finding solutions for the limitations of the approach, extending PRO-TEST to existing diagrams, and other non-functional requirements. We have identified the following possible directions for future work, which would be of benefit to researchers who are interested in this area.

\begin{enumerate}
    \item Apply the proposed modeling technique on a larger set of well-built SRS with relatively completed performance requirements and to enhance PRO-TEST further.
    \item Investigate the possibility of implementing the relational modeling concept in other non-functional requirements, e.g., security.
    \item Integrate PRO-TEST with MBT approaches that generate functional test cases, and evaluate the effectiveness of test environment generation.
    \item Extend the taxonomy tree by finding the possible sub-categories for the performance aspects.
    \item Automate the process of creating the model from natural language requirements to avoid human errors.
\end{enumerate}

Finally, we hope that this list of future work inspires researchers to do more research in the area of model-based performance testing and performance requirements veri.

\bibliographystyle{spmpsci}
\bibliography{references.bib}

\appendix
\section{Included Papers in the SMS}
\label{app:included-papers-in-sms}

\onecolumn
    \begin{longtable}{p{0.05\textwidth}p{0.64\textwidth}p{0.16\textwidth}p{0.05\textwidth}}
    \caption{Included papers in the SMS}
\label{tab:Included papers in the SMS}
        \\
        \hline\noalign{\smallskip}
        \textbf{No.} 
        &
        \textbf{Title} 
        &
        \textbf{Author}
        &
        \textbf{Year}
        \\
        \noalign{\smallskip}\hline\noalign{\smallskip}
        S1 
        &
        Model-based performance testing in the cloud using the mbpet tool
        &
        Abbors et al.
        &
        2013
        \\
        S2
        &
        Approaching performance testing from a model-based testing perspective
        &
        Abbors et al.
        &
        2010
        \\
        S3
        &
        Model-based testing of a real-time adaptive motion planning system
        &
        Abdelgawad et al.
        &
        2017
        \\
        S4
        &
        GeTeX: A Tool for Testing Real-Time Embedded Systems Using CAN Applications
        &
        AbouTrab et al.
        &
        2011
        \\
        S5
        &
        Test generation for performance evaluation of mobile multimedia streaming applications
        &
        Al-tekreeti et al.
        &
        2018
        \\
        S6
        &
        Dtron: a tool for distributed model-based testing of time critical applications
        &
        Anier et al.
        &
        2017
        \\
        S7
        &
        Canopus: A Domain-Specific Language for Modeling Performance Testing
        &
        Bernardino et al.
        &
        2016
        \\
        S8
        &
        Online model-based testing under uncertaint
        &
        Camilli et al.
        &
        2018
        \\
        S9
        &
        Event-based runtime verification of temporal properties using time basic Petri nets
        &
        Camilli et al.
        &
        2017
        \\
        S10
        &
        Abstracting timing information in UML state charts via temporal ordering and LOTOS
        &
        Chimisliu et al.
        &
        2011
        \\
        S11
        &
        Generation of scripts for performance testing based on UML models
        &
        Da Silveira et al.
        &
        2011
        \\
        S12
        &
        Timed testing under partial observability
        &
        David et al.
        &
        2009
        \\
        S13
        &
        Model-Based Test Suite Generation for Function Block Diagrams Using the UPPAAL Model Checker
        &
        Enoiu et al.
        &
        2013
        \\
        S14
        &
        Iterative test suites refinement for elastic computing systems
        &
        Gambi et al.
        &
        2013
        \\
        S15
        &
        Fast model-based test case classification for performance analysis of multimedia mpsoc platforms
        &
        Gangadharan et al.
        &
        2009
        \\
        S16
        &
        Fault-driven stress testing of distributed real-time software based on uml models
        &
        Garousi
        &
        2011
        \\
        S17
        &
        Automated Steering of Model-Based Test Oracles to Admit Real Program Behaviors
        &
        Gay et al.
        &
        2011
        \\
        S18
        &
        Model-driven testing approach for embedded systems specifics verification based on UML model transformation
        &
        Grigorjevs
        &
        2011
        \\
        S19
        &
        Usage profile and platform independent automated validation of service behavior specifications
        &
        Groenda
        &
        2010
        \\
        S20
        &
        A model-based testing technique for component-based real-time embedded systems
        &
        Guan et al.
        &
        2015
        \\
        S21
        &
        Validating Timed Component Contracts
        &
        Guilly et al.
        &
        2015
        \\
        S22
        &
        Towards effective and scalable testing for complex high-speed railway signal software
        &
        Hu et al.
        &
        2017
        \\
        S23
        &
        Experiences of Applying UML/MARTE on Three Industrial Projects
        &
        Iqbal et al.
        &
        2012
        \\
        S24
        &
        Environment modeling and simulation for automated testing of soft real-time embedded software
        &
        Iqbal et al.
        &
        2015
        \\
        S25
        &
        Applicability of an integrated model-based testing approach for rtes
        &
        Iyenghar et al.
        &
        2011
        \\
        S26
        &
        Model-Driven Method for Performance Testing
        &
        Javed et al.
        &
        2018
        \\
        S27
        &
        Experience Report: Evaluating fault detection effectiveness and resource efficiency of the architecture quality assurance framework and tool
        &
        Johnsen et al.
        &
        2017
        \\
        S28
        &
        Interaction-based runtime verification for systems of systems integration
        &
        Krüger et al.
        &
        2010
        \\
        S29
        &
        Quality Assurance for Component-based Systems in Embedded Environments
        &
        Li et al.
        &
        2018
        \\
        S30
        &
        Timed moore automata: test data generation and model checking
        &
        Löding et al.
        &
        2010
        \\
        S31
        &
        Minimum/maximum delay testing of product lines with unbounded parametric real-time constraints.
        &
        Luthmann et al.
        &
        2019
        \\
        S32
        &
        Modeling and testing product lines with unbounded parametric real-time constraints
        &
        Luthmann et al.
        &
        2017
        \\
        S33
        &
        Automated significant load testing for ws-bpel compositions
        &
        Maâlej et al.
        &
        2013
        \\
        S34
        &
        Conformance testing for quality assurance of clustering architectures
        &
        Maâlej et al.
        &
        2013
        \\
        S35
        &
        Model-based conformance testing of ws-bpel compositions
        &
        Maâlej et al.
        &
        2012
        \\
        S36
        &
        Towards an industrial strength process for timed testing
        &
        Mitsching et al.
        &
        2009
        \\
        S37
        &
        Comparative analysis for software testing: Mobile applications versus web applications
        &
        Muhamad et al.
        &
        2016
        \\
        S38
        &
        Test Selection for Data-Flow Reactive Systems Based on Observations
        &
        Nguena-Timo et al.
        &
        2011
        \\
        S39
        &
        PLeTsPerf - A Model-Based Performance Testing Tool
        &
        Rodrigues et al.
        &
        2015
        \\
        S40
        &
        Evaluating capture and replay and model-based performance testing tools: an empirical comparison
        &
        Rodrigues et al.
        &
        2014
        \\
        S41
        &
        Extending UML testing profile towards non-functional test modeling
        &
        Rodrigues et al.
        &
        2014
        \\
        S42
        &
        An experience report on an industrial case-study about timed model-based testing with UPPAAL-TRON
        &
        Rütz et al.
        &
        2011
        \\
        S43
        &
        Testing of timing properties in real-time systems: Verifying clock constraints
        &
        Saadatmand et al.
        &
        2013
        \\
        S44
        &
        On Combining Model-Based Analysis and Testing
        &
        Saadatmand et al.
        &
        2013
        \\
        S45
        &
        Functionality, performance, and compatibility testing: A model based approach
        &
        Saqib et al.
        &
        2018
        \\
        S46
        &
        Checking response-time properties of web-service applications under stochastic user profiles
        &
        Schumi et al.
        &
        2017
        \\
        S47
        &
        Analyzing a wind turbine system: From simulation to formal verification
        &
        Seceleanu et al.
        &
        2017
        \\
        S48
        &
        Introduction of time and timing variability in usage model based testing
        &
        Siegl et al.
        &
        2010
        \\
        S49
        &
        Partitioning the requirements of embedded systems by input/output dependency analysis for compositional creation of parallel test models
        &
        Siegl et al.
        &
        2015
        \\
        S50
        &
        Multi-fragment Markov model guided online test generation for MPSoC
        &
        Vain et al.
        &
        2017
        \\
        S51
        &
        Provably Correct Test Development for Timed Systems
        &
        Vain et al.
        &
        2014
        \\
        S52
        &
        System Testing of Timing Requirements Based on Use Cases and Timed Automata
        &
        Wang et al.
        &
        2017
        \\
        S53
        &
        A model-based framework for cloud api testing
        &
        Wang et al.
        &
        2017
        \\
        S54
        &
        Towards an integrated approach for validating qualities of self-adaptive systems
        &
        Weyns
        &
        2012
        \\
        S55
        &
        Vision paper: Towards model-based energy testing
        &
        Wilke et al.
        &
        2011
        \\
        S56
        &
        System Modules Interaction Based Stress Testing Model
        &
        Yang et al.
        &
        2010
        \\
        S57
        &
        A methodology of model-based testing for aadl flow latency in cps
        &
        Zhu et al.
        &
        2011
        \\
        \noalign{\smallskip}\hline
    \end{longtable}
    \twocolumn

\onecolumn
    \begin{longtable}{p{0.05\textwidth}p{0.64\textwidth}p{0.16\textwidth}p{0.05\textwidth}}
    \caption{Extracted Papers from Dias-Neto 2010}
\label{tab:Papers exported from dias neto}

        \\
        \hline\noalign{\smallskip}
        \textbf{No.} 
        &
        \textbf{Title} 
        &
        \textbf{Author}
        &
        \textbf{Year}
        \\
        \noalign{\smallskip}\hline\noalign{\smallskip}
        S58
        &
        Specification-based testing for real-time reactive systems
        &
        Alagar et al.
        &
        2000
        \\
        S59
        &
        Designing fault injection experiments using state-based model to test a space software
        &
        Ambrosio et al.
        &
        2007
        \\
        S60
        &
        Generating test suites for software load testing
        &
        Avritzer et al.
        &
        1994
        \\
        S61
        &
        Specification-based testing for real-time avionic systems
        &
        Biberstein et al.
        &
        1999
        \\
        S62
        &
        On the correctness of upper layers of automotive systems
        &
        Botaschanjan et al.
        &
        2008
        \\
        S63
        &
        Distributed software testing with specification
        &
        Chang et al.
        &
        1990
        \\
        S64
        &
        Traffic-aware stress testing of distributed systems based on UML models
        &
        Garousi et al.
        &
        2006
        \\
        S65
        &
        Testing from a stochastic timed system with a fault model
        &
        Hierons et al.
        &
        2009
        \\
        S66
        &
        Automatic timed test case generation for Web services composition
        &
        Lallali et al. 
        &
        2008
        \\
        S67
        &
        Regression testing of classes based on TCOZ specification
        &
        Liang
        &
        2005
        \\
        S68
        &
        Generating test cases for real-time systems from logic specifications
        &
        Mandrioli et al.
        &
        1995
        \\
        S69
        &
        Derivation of tests from timed specifications according to different coverage criteria
        &
        Merayo et al.
        &
        2008
        \\
        S70
        &
        T-UPPAAL: online model-based testing of real-time systems
        &
        Mikucionis et al.
        &
        2004
        \\
        S71
        &
        Generating functional test cases in-the-large for time-critical systems from logic-based specifications
        &
        Morasca et al.
        &
        1996
        \\
        S72
        &
        Mutation-based Testing Criteria for Timeliness
        &
        Nilson et al.
        &
        2004
        \\
        S73
        &
        Model-based testing in evolutionary software development
        &
        Pretschner et al.
        &
        2001
        \\
        S74
        &
        Specification-based test oracles for reactive systems
        &
        Richardson et al.
        &
        1992
        \\
        S75
        &
        Model-based testing of object-oriented systems
        &
        Rumpe
        &
        2003
        \\
        S76
        &
        Aiding modular design and verification of safety-critical time-triggered systems by use of executable formal specifications
        &
        Sakurai et al.
        &
        2008
        \\
        S77
        &
        An evaluation of a model-based testing method for information systems
        &
        Santos-Neto et al.
        &
        2008
        \\
        \noalign{\smallskip}\hline
    \end{longtable}
    \twocolumn

\end{document}